%% file: ICA2022_ABS-0237.tex
\documentclass[10.5pt,a4paper,oneside]{extarticle}
\usepackage[T1]{fontenc} 
\usepackage[left=25mm,right=20.9mm,top=4.25cm,bottom=3.75cm,headsep=0.25cm,footskip=0.25cm]{geometry}

\usepackage[english]{babel}
    \usepackage[affil-it]{authblk}
    \usepackage{etoolbox}
    \usepackage{lmodern}
    \usepackage{setspace}
    \usepackage{amssymb,amsthm,amsmath,bm}
    \usepackage{tikz}
    \usepackage{listings}
    \usepackage{color}
    \usepackage{caption}
    \usepackage{placeins}
    \usepackage{booktabs}
    \usepackage{enumerate}
    \usepackage{graphicx}
    \usepackage{titlesec}
    \usepackage{mathptmx}
    \usepackage{anyfontsize}
    \usepackage{t1enc}
    \usepackage[utf8]{inputenc}
    \usepackage{fancyhdr}
    \usepackage{float,xspace}
    \usepackage{scalefnt}
    \usepackage{siunitx}
    \usepackage{tabularx}
    \usepackage{multirow}
    \usepackage{subcaption}
    \usepackage{pgfplots}
    \usetikzlibrary{plotmarks}
    \usetikzlibrary{calc}

    \usetikzlibrary{arrows.meta}
    \usepgfplotslibrary{patchplots}
    \usepackage{grffile}
    \usepackage{nicefrac}
    \usepackage{xfrac}

\pgfplotsset{compat=1.16}
\usetikzlibrary{decorations.markings}
\usetikzlibrary{patterns}

\newenvironment{customlegend}[1][]{%
    \begingroup
    \csname pgfplots@init@cleared@structures\endcsname
    \pgfplotsset{#1}%
}{%
    \csname pgfplots@createlegend\endcsname
    \endgroup
}%
\def\addlegendimage{\csname pgfplots@addlegendimage\endcsname}

\DeclareInstance{xfrac}{\familydefault}{math}
{
    slash-symbol-font=ptm
}

\DeclareMathOperator*{\argi}{arg\,min}

\def\argmin#1#2{\argi_{\substack{#1}}\;\left\Vert #2 \right\Vert}

\def\*#1{\mathbf{#1}} 

\usepackage{soul} 
\usepackage{tabulary}
\usepackage[colorlinks=true,linkcolor=blue]{hyperref} 

\usepackage{textcase}

\usepackage[colorinlistoftodos]{todonotes}

\titlespacing*{\section}
{0pt}{12pt}{0pt}
\titlespacing*{\subsection}
{0pt}{12pt}{0pt}
\titlespacing*{\subsubsection}
{0pt}{12pt}{0pt}
\titleformat{\section}
  {\huge\sffamily\Large\bfseries\color{black}}
  {\thesection}{1em}{}
    \titleformat{\subsection}[block]{\bfseries}{\thesubsection}{.5em}{}
    \titleformat{\subsubsection}[block]{\bfseries}{\thesubsubsection}{.5em}{}

\titleformat{\section}{\fontsize{12}{19}\bfseries}{\thesection}{1em}{}

\usepackage{mathptmx} 

\makeatletter

\patchcmd{\@maketitle}{\LARGE \@title}{\fontsize{14}{19.2}\selectfont\@title}{}{} 
\makeatother
\title
{
	\vspace{-2cm}
 \textbf{Speech enhancement using ego-noise references with a microphone array embedded in an unmanned aerial vehicle}
	\author[ ]{Elisa TENGAN$^{(1)}$, Thomas DIETZEN$^{(1)}$, Santiago RUIZ$^{(1)}$, Mansour ALKMIM$^{(2,3)}$, João CARDENUTO$^{(2,3)}$, Toon VAN WATERSCHOOT$^{(1)}$}
  	\affil[(1)]{Dept. of Electrical Engineering (ESAT-STADIUS), KU Leuven, Leuven, Belgium, elisa.tengan@esat.kuleuven.be}
  	\affil[(2)]{Siemens Digital Industries Software, Leuven, Belgium}
  	\affil[(3)]{Dept. of Mechanical Engineering, KU Leuven, Leuven, Belgium}
}
\date{}

\begin{document}

\clearpage
\setcounter{page}{1}
\maketitle
\thispagestyle{empty}
\fancypagestyle{empty}
{
	\fancyhf{} \fancyfoot[R]
	{
	}
}

\subsection*{\fontsize{10.5}{19.2}\uppercase{\textbf{Abstract}}}
{\fontsize{10.5}{60}\selectfont A method is proposed for performing speech enhancement using ego-noise references with a microphone array embedded in an unmanned aerial vehicle (UAV). The ego-noise reference signals are captured with microphones located near the UAV’s propellers and used in the prior knowledge multichannel Wiener filter (PK-MWF) to obtain the speech correlation matrix estimate. Speech presence probability (SPP) can be estimated for detecting speech activity from an external microphone near the speech source, providing a performance benchmark, or from one of the embedded microphones, assuming a more realistic scenario. Experimental measurements are performed in a semi-anechoic chamber, with a UAV mounted on a stand and a loudspeaker playing a speech signal, while setting three distinct and fixed propeller rotation speeds, resulting in three different signal-to-noise ratios (SNRs). The recordings obtained and made available online are used to compare the proposed method to the use of the standard multichannel Wiener filter (MWF) estimated with and without the propellers’ microphones being used in its formulation. Results show that compared to those, the use of PK-MWF achieves higher levels of improvement in speech intelligibility and quality, measured by STOI and PESQ, while the SNR improvement is similar.}

\noindent{\\ \fontsize{11}{60}\selectfont Keywords: Speech enhancement, unmanned aerial vehicle, noise reduction} 

\fontdimen2\font=4pt

\section{\uppercase{Introduction}}
The use of unmanned aerial vehicles (UAVs), usually referred to as drones, has not only become more common in modern society, but also more diverse in terms of its applications and consequently, the digital signal processing solutions explored.
In situations where drones are used for cinematography, surveillance and emergency search and rescue operations, the potential of recording and processing audio signals, as opposed to limiting the use of UAVs to image and video capture only, has become more evident and researched over the past few years \cite{dregon, hioka_speechenhancement}. A fundamental problem in recording audio with a UAV, however, is the high level of ego-noise generated by its rotors \cite{nasa, book_hioka}, which interferes with the target source signal and consequently reduces sound quality. In order to overcome this issue, noise reduction and speech enhancement techniques can be employed \cite{gannot}.

In the current state of the art, noise reduction and speech enhancement frameworks based on the standard multichannel Wiener filter (MWF), computed as a function of the speech source steering vector, or on blind source separation (BSS) methods, have already been used and tested on UAV setups \cite{hioka_speechenhancement, cavallaro}. However, these approaches require the speech source location to be presumably known, which can be a limiting factor in more practical scenarios. An alternative formulation of the standard MWF allows the filter coefficients to be computed as a function of the speech correlation matrix instead \cite{moonen}, which is in turn estimated by using speech activity information obtained with a voice activity detector (VAD) or with speech presence probability (SPP) estimates \cite{ngo}. When considering this formulation, an extension of the standard MWF can be realized when prior knowledge on the noise is available from specific channels of the microphone array and used to improve robustness in the estimation of the speech correlation matrix, resulting in the prior knowledge multichannel Wiener filter (PK-MWF) \cite{rompaey}.

In this paper, a method is proposed for performing speech enhancement using ego-noise references with a microphone array embedded in a UAV, while assuming that the source location, and consequently, the source steering vector with respect to the array configuration, is unknown. The ego-noise reference signals are captured with microphones located near the UAV’s propellers and used to constrain the estimation of the speech correlation matrix necessary for computing the prior knowledge multichannel Wiener filter (PK-MWF). Speech presence probability (SPP) information is estimated in order to identify speech activity in different time-frequency points. Such SPP estimation can either be obtained from one of the embedded array's microphones or from an external microphone also used for recording the auditory scene and providing a performance benchmark. Experimental measurements, made available online \cite{github}, were performed in a semi-anechoic chamber for testing the implementation of the standard MWF and the PK-MWF. It will be shown that the use of PK-MWF achieves higher levels of improvement than the standard MWF in terms of short-time objective intelligibility (STOI) and perceptual evaluation of speech quality (PESQ), while the SNR improvement is similar.

The paper is organized as follows. In section \ref{sec:signalmodel}, the signal model is defined. In section \ref{sec:method}, the proposed method is explained and its formulation is compared with the standard multichannel Wiener filter (MWF). In section \ref{sec:evaluation}, the experimental setup used for the recordings and the audio processing framework are described. In section \ref{sec:results}, the results obtained are discussed and finally, in section \ref{sec:conclusion}, a conclusion is presented with a summary and final remarks on the work accomplished.

\section{\uppercase{Signal model}}
\label{sec:signalmodel}
The signal in the short-time Fourier transform (STFT) domain captured by a single microphone with index $m$ is defined as
\begin{align}
  y_m\left(\kappa,l\right) &= a_m\left(\kappa,l\right)s\left(\kappa,l\right)+n_m\left(\kappa,l\right),
\end{align}
where $y_m$ is the resulting signal, $a_m$ is the corresponding transfer function from the target source to the microphone, $s$ is the speech source signal, and $n_m$ is the noise component. The indices $\kappa$ and $l$ correspond to the frequency bin index and the observation frame index, respectively. In the following equations, we consider processing the different frames and frequency bins independently, and their corresponding indices will be omitted for brevity.
Considering a microphone array with $M$ elements, a signal vector is obtained by stacking all microphone signal equations, resulting in
\begin{align}
  \mathbf{y} &= \mathbf{s}+\mathbf{n}\\
&=\mathbf{a}s+\mathbf{n},\label{eq:sigmodel}
\end{align}
with $\left\{\mathbf{y},\mathbf{s},\mathbf{a},\mathbf{n} \right\} \in \mathbb{C}^{M}$, and $s \in \mathbb{C}$.
Assuming that the desired target source signal and the noise component are uncorrelated, the correlation matrices adhere to
\begin{align}
  \mathbf{R_{yy}} &= \mathbb{E}\left\{\mathbf{y}\mathbf{y}^{\mathrm{H}}\right\}\\
  & = \mathbb{E}\left\{\mathbf{s}\mathbf{s}^{\mathrm{H}}\right\} + \mathbb{E}\left\{\mathbf{n}\mathbf{n}^{\mathrm{H}}\right\}\\
  &= \mathbf{R_{ss}}+\mathbf{R_{nn}},\label{eq:corrmat}
\end{align}
where $\mathbb{E}\{.\}$ denotes the expectation operator, $\left(.\right)^{\mathrm{H}}$ denotes the conjugate transpose, $\mathbf{R_{yy}}$ is the microphone signal correlation matrix, $\mathbf{R_{ss}}$ and $\mathbf{R_{nn}}$ are the speech signal and noise correlation matrices, respectively. The speech correlation matrix corresponds to $\mathbf{R_{ss}}=\mathbf{a}\mathbf{a}^{\mathrm{H}}\mathbb{E}\left\{s^2\right\}$,
and hence it is rank-1 under the common assumption that $\mathbf{a}$ is deterministic.

\section{\uppercase{Proposed method}}
\label{sec:method}
We propose to use the PK-MWF with prior knowledge obtained from microphones mounted below the rotors of the UAV. In section \ref{sec:method:mwf}, we introduce the general formulation of the multi-channel Wiener filter. In section \ref{sec:method:mwf_realization}, we outline the standard MWF realization, and in section \ref{sec:method:pkmwf_realization}, we describe the PK-MWF realization and its application in the context of a UAV.

\subsection{Formulation of the multichannel Wiener filter (MWF)}
\label{sec:method:mwf}
Let a target speech reference signal $d$ be defined as
\begin{align}
  d = \mathbf{e}_d^{\mathrm{T}}\mathbf{s},\label{eq:targetd1}
\end{align}
with $\mathbf{e}_d = [1, 0, \hdots, 0]^{\mathrm{T}}$ being a vector selecting the desired source signal component in the microphone array's first channel, which is here without loss of generality considered as the reference channel, and $(.)^{\mathrm{T}}$ denoting the transpose.

In the formulation of the multichannel Wiener filter (MWF), it is aimed to estimate $d$ as a linear combination of all microphone signals in $\mathbf{y}$ by minimizing the following mean squared error (MSE) criterion and obtaining the filter weights $\bar{\mathbf{w}}$ as \cite{moonen}
\begin{align}
\bar{\mathbf{w}} = \argi_{\mathbf{w}} \mathbb{E}\left\{|d - \mathbf{w}^{\mathrm{H}}\mathbf{y}|^2\right\}.\label{eq:MWF_costfct1}
\end{align}
If the microphone signal correlation matrix $\mathbf{R_{yy}}$ has full rank, the optimal solution to (\ref{eq:MWF_costfct1}) is \cite{moonen}
\begin{align}
  \bar{\mathbf{w}} = \mathbf{R_{yy}^{-1}}\mathbf{R_{ss}}\mathbf{e}_d\label{eq:MWF_sol1}.
\end{align}
Since the speech-only correlation matrix $\mathbf{R_{ss}}$ cannot be directly observed from the microphone signals due to the presence of noise, its estimate can be based on the analysis of the speech activity's on-off behavior in the time-frequency domain. While assuming short-term stationarity of $\mathbf{y}$, the estimation of the speech-plus-noise and noise-only correlation matrices ($\mathbf{\hat{R}_{yy}}$ and $\mathbf{\hat{R}_{nn}}$, respectively) can be performed instead and used to estimate $\mathbf{\hat{R}_{ss}}$.

\subsection{Standard multichannel Wiener filter (MWF) realization}
\label{sec:method:mwf_realization}
In the standard multichannel Wiener filter, the speech-only correlation matrix is estimated by solving the following optimization problem \cite{serizel}:
\begin{align}
\mathbf{\hat{R}_{ss}} = \argmin{\mathrm{rank}(\*{R_{ss}})=1 \\ \*{R_{ss}} \succeq 0} {\*{\hat{R}^\mathrm{-\frac{1}{2}}_{nn}} \left( \*{\hat{R}_{yy} - \hat{R}_{nn} - R_{ss}}  \right) \*{\hat{R}}^{-\mathrm{\frac{H}{2}}}_{\*{nn}}}^2_\mathrm{F},
 \label{eq:argminRss_noPK}
\end{align}
where $\|.\|_\mathrm{F}$ denotes the Frobenius norm and $\mathbf{\hat{R}_{ss}}$ is constrained to be a rank-1 and positive semi-definite matrix. The solution to \eqref{eq:argminRss_noPK} is based on the generalized eigenvalue decomposition (GEVD) of the matrix pencil $\{\mathbf{\hat{R}_{yy}}$, $\mathbf{\hat{R}_{nn}}\}$ \cite{serizel}:
\begin{align}
  \hat{\mathbf{R}}_{\mathbf{yy}} &= \hat{\mathbf{Q}}\hat{\bm{\Sigma}}_{\mathbf{yy}}\hat{\mathbf{Q}}^\mathrm{H},\label{eq:Ryy}\\
  \hat{\mathbf{R}}_{\mathbf{nn}} &= \hat{\mathbf{Q}}\hat{\bm{\Sigma}}_{\mathbf{nn}}\hat{\mathbf{Q}}^\mathrm{H},\label{eq:Rnn}
\end{align}
where $\hat{\bm{\Sigma}}_{\mathbf{yy}}$ and $\hat{\bm{\Sigma}}_{\mathbf{nn}}$ are diagonal matrices containing the generalized eigenvalues $\hat{\sigma}_{y_i}$ and $\hat{\sigma}_{n_i}$ of $\hat{\mathbf{R}}_{\mathbf{yy}}$ and $\hat{\mathbf{R}}_{\mathbf{nn}}$,
 respectively, and $\hat{\mathbf{Q}}$ is an invertible matrix containing the generalized eigenvectors in its columns. The generalized eigenvalues and eigenvectors are assumed to be sorted in descending order of $\hat{\sigma}_{y_i}$ and $\hat{\sigma}_{n_i}$.
Using \eqref{eq:corrmat}, the estimate of $\mathbf{\hat{R}_{ss}}$ can then be obtained as
\begin{align}
  \mathbf{\hat{R}_{ss}} = \hat{\mathbf{Q}}\,\text{diag}\{ \hat{\sigma}_{y_1} - \hat{\sigma}_{n_1} ,0,\hdots,0\}\hat{\mathbf{Q}}^\mathrm{H},\label{eq:Rss}
\end{align}
with $\text{diag}\{.\}$ building a diagonal matrix and $\hat{\sigma}_{y_1}$ and $\hat{\sigma}_{n_1}$ being the top-left elements of $\hat{\bm{\Sigma}}_{\mathbf{yy}}$ and $\hat{\bm{\Sigma}}_{\mathbf{nn}}$, respectively, yielding the largest ratio between eigenvalues $\hat{\sigma}_{y_i}/\hat{\sigma}_{n_i}$. Finally, by substituting (\ref{eq:Rss}) and (\ref{eq:Ryy}) into (\ref{eq:MWF_sol1}), the standard multichannel Wiener filter can be formulated as \cite{serizel}
\begin{align}
  \hat{\mathbf{w}}_{\text{MWF}} = \hat{\mathbf{Q}}^{-\mathrm{H}}\text{diag}\left\{1 - \frac{\hat{\sigma}_{n1}}{\hat{\sigma}_{y1}},0,\hdots,0\right\}\hat{\mathbf{Q}}^{\mathrm{H}}\mathbf{e}_d.
\end{align}

\subsection{Prior knowledge multichannel Wiener filter (PK-MWF) realization}
\label{sec:method:pkmwf_realization}
In the standard multichannel Wiener filter formulation, the spatial configuration of the microphone array used with respect to the noise sources is not explicitly exploited. However, for a practical setup such as the microphone array embedded in a UAV considered in this study, the prior knowledge obtained from array elements sufficiently close to the main noise sources interfering in the signals, i.e. the device's rotors, can be used in an attempt to provide a more robust estimation of $\mathbf{R_{ss}}$.

We define $M_{\text{S+N}}$ as the number of array elements that are assumed to contain speech and noise, and $M_{\text{N}}$ as the number of array elements assumed to only contain noise, or have a sufficiently low signal-to-noise ratio (SNR) such that the speech component is negligible. In a UAV setup, such configuration is here assumed to be attained if each of the $M_{\text{N}}$ microphones are placed in close proximity to one of the vehicle's propellers. The total number of microphones used are then described as $M=M_{\text{S+N}}+M_{\text{N}}$.
Let the identity and all-zero matrix be denoted by $\mathbf{I}$ and $\mathbf{0}$, respectively, where we indicate their dimensions by a subscript. We define the selection matrix $\mathbf{H}$ and the blocking matrix $\mathbf{B}$ as
\begin{align}
  \mathbf{H} = \begin{bmatrix}
\mathbf{I}_{M_{\text{S+N}}}\\
\mathbf{0}_{M_{\text{N}}\times M_{\text{S+N}}}
\end{bmatrix},
\qquad \mathbf{B} = \begin{bmatrix}
\mathbf{0}_{M_{\text{S+N}}\times M_{\text{N}}}\\
\mathbf{I}_{M_{\text{N}}}
\end{bmatrix}.
\end{align}
Thus, the selection matrix $\mathbf{H}$ and blocking matrix $\mathbf{B}$ have dimensions $M\times M_{\text{S+N}}$ and $M\times M_{\text{N}}$, respectively.
In the PK-MWF, the following cost function is then minimized in order to estimate $\mathbf{\hat{R}_{ss}}$ \cite{rompaey}:
\begin{align}
  \*{\hat{R}_{ss}} =\argmin{\mathrm{rank}(\*{R_{ss}})=1 \\ \*B^{\mathrm{H}} \*{R_{ss}B}=\mathbf{0} \\ \*{R_{ss}} \succeq 0} {\*{\hat{R}^{-\mathrm{\frac{1}{2}}}_{nn}} \left( \*{\hat{R}_{yy} - \hat{R}_{nn} - R_{ss}} \right) \*{\hat{R}}^{-\mathrm{\frac{H}{2}}}_{\*{nn}}}^2_\mathrm{F}.
\label{eq:PKMWF_Rss}
\end{align}
The matrix $\mathbf{\hat{R}_{ss}}$ to be estimated now, which is the counterpart to \eqref{eq:argminRss_noPK} in the standard MWF, is not only constrained to be rank-1 and positive semi-definite, but also to have its column and row spaces lying in the column space of $\mathbf{H}$, such that $\*B^{\mathrm{H}} \*{R_{ss}B}=\mathbf{0}$.
The solution to \eqref{eq:PKMWF_Rss} (proof omitted) is obtained by firstly applying a linearly-constrained minimum variance (LCMV) beamformer $\mathbf{C}$ to the vector $\mathbf{y}$, defined by the following criterion \cite{rompaey}:
\begin{align}
  \mathbf{C} = \argi_{\mathbf{C}} \quad&{\text{trace\,}\left\{ \mathbf{C}^{\mathrm{H}}\mathbf{R_{nn}}\mathbf{C}\right\}}\\
  \text{s.t.\quad}&\mathbf{H}^{\mathrm{H}}\mathbf{C}=\mathbf{I}_{M_{\text{S+N}}},
\end{align}
which has its solution based on a generalized sidelobe canceler (GSC) \cite{strauss}:
\begin{align}
  \hat{\mathbf{C}} &= \mathbf{H}-\mathbf{B}\hat{\mathbf{F}},\label{eq:GSC_C}\\
    \hat{\mathbf{F}} &= \left(\mathbf{B}^{\mathrm{H}}\hat{\mathbf{R}}_{\mathbf{nn}}\mathbf{B}\right)^{-1}\mathbf{B}^{\mathrm{H}} \hat{\mathbf{R}}_{\mathbf{nn}} \mathbf{H}.\label{eq:GSC_F}
\end{align}
By applying $\hat{\mathbf{C}}$ to $\mathbf{y}$, we obtain a vector with reduced dimension $\mathbf{y}^{\text{red}}=\hat{\mathbf{C}}^{\mathrm{H}}\mathbf{y}$. Similarly, we can obtain the reduced dimension correlation matrices $\hat{\mathbf{R}}^{\text{red}}_{\mathbf{yy}} = \hat{\mathbf{C}}^{\mathrm{H}}\hat{\mathbf{R}}_{\mathbf{yy}}\hat{\mathbf{C}}$
and
$\hat{\mathbf{R}}^{\text{red}}_{\mathbf{nn}} = \hat{\mathbf{C}}^{\mathrm{H}}\hat{\mathbf{R}}_{\mathbf{nn}}\hat{\mathbf{C}}$.
Then, a generalized eigenvalue decomposition (GEVD) of the reduced matrix pencil $\left\{\hat{\mathbf{R}}^{\text{red}}_{\mathbf{yy}},\hat{\mathbf{R}}^{\text{red}}_{\mathbf{nn}}\right\}$
is performed as in \cite{rompaey}
\begin{align}
  \hat{\mathbf{R}}^{\text{red}}_{\mathbf{yy}} &= \hat{\mathbf{Q}}^{\text{red}}\bm{\hat{\Sigma}}^{\text{red}}_{\mathbf{yy}}(\hat{\mathbf{Q}}^{\text{red}})^{\mathrm{H}}\label{eq:gevd_Ryy1},\\
    \hat{\mathbf{R}}^{\text{red}}_{\mathbf{nn}} &= \hat{\mathbf{Q}}^{\text{red}}\hat{\bm{\Sigma}}_{\mathbf{nn}}^{\text{red}}(\hat{\mathbf{Q}}^{\text{red}})^{\mathrm{H}}\label{eq:gevd_Rnn1},
\end{align}
which is the PK-MWF counterpart to \eqref{eq:Ryy}-\eqref{eq:Rnn} in the standard MWF. Again, we assume the generalized eigenvalues and eigenvectors to be sorted in descending order of $\hat{\sigma}_{y_i}^{\text{red}}$ and $\hat{\sigma}_{n_i}^{\text{red}}$. The matrix $\mathbf{\hat{R}_{ss}}$ can then be expressed as \cite{rompaey}
\begin{align}
  \hat{\mathbf{R}}_{\mathbf{ss}} = \mathbf{H}\hat{\mathbf{Q}} ^{\text{red}}\ \text{diag}\left\{\hat{\sigma}_{y1}^{\text{red}} - \hat{\sigma}_{n1}^{\text{red}},0,...,0\right\}(\hat{\mathbf{Q}}^{\text{red}})^{\mathrm{H}}\mathbf{H}^{\mathrm{H}},
  \label{eq:Rss_pkmwf}
\end{align}
which corresponds to the PK-MWF counterpart to \eqref{eq:Rss} in the standard MWF. Finally, by substituting \eqref{eq:Rss_pkmwf} and \eqref{eq:gevd_Ryy1} into \eqref{eq:MWF_sol1}, the estimated prior knowledge multichannel Wiener filter (PK-MWF) coefficients are given by
\begin{align}
  \hat{\mathbf{w}}_{\text{PK-MWF}} = \hat{\mathbf{C}}(\hat{\mathbf{Q}}^{\text{red}})^{\mathrm{-H}}\text{diag}\left\{1 - \frac{\hat{\sigma}_{n1}^{\text{red}}}{\hat{\sigma}_{y1}^{\text{red}}},0,\hdots,0\right\}(\hat{\mathbf{Q}}^{\text{red}})^{\mathrm{H}}\mathbf{H}^{\mathrm{H}}\mathbf{e}_d.
  \label{eq:PKMWF_hat}
\end{align}

\section{\uppercase{Evaluation}}
\label{sec:evaluation}
In order to compare the performance of both multichannel filtering methods specified in section \ref{sec:method}, experimental tests were carried out in a semi-anechoic chamber at Siemens Digital Industries Software, in Leuven, Belgium. In section \ref{sec:setup}, the measurement setup is detailed, and in section \ref{sec:processing}, the processing framework and the different cases studied are described.
\subsection{Measurement setup}
\label{sec:setup}
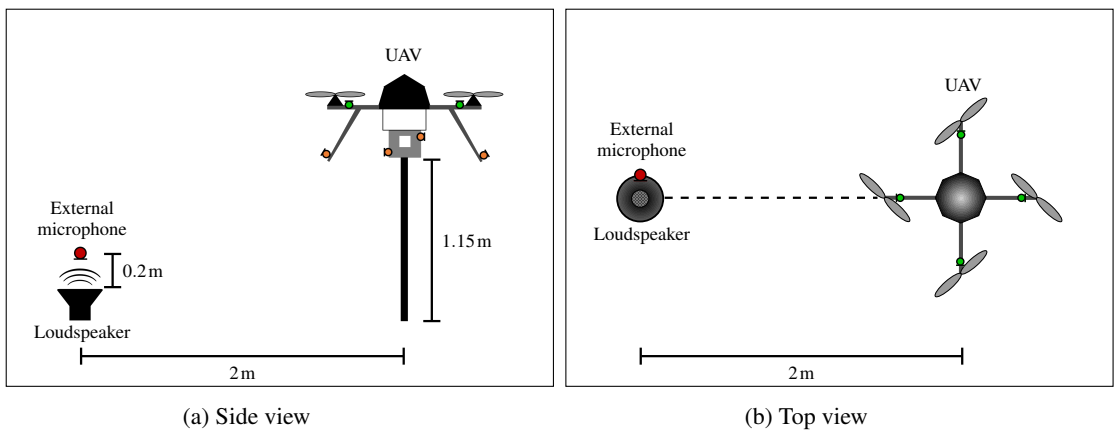
\begin{figure}[h!]
  \centering
\begin{subfigure}{.4\textwidth}
  \centering
  \tikzset{every picture/.style={line width=0.75pt}} 
  \begin{tikzpicture}
  \pgfdeclareimage[width=.9\textwidth]{sketch}{./figures/DRONE_CONFERENCE_SKETCH1}
  \node (img1) at (0,0) {\pgfuseimage{sketch}};
  \draw[|-|,line width=0.3mm] (-2.63,-2.1) -- node[below] {\scriptsize $2\,\text{m}$} ++ (4.275,0);
  \draw[|-|,line width=0.3mm] (2.,-1.65) -- node[right] {\scriptsize $1.15\,\text{m}$} ++ (0,2.17);
  \draw[|-|,line width=0.3mm] (-2.2,-1.2) -- node[right] {\scriptsize $0.2\,\text{m}$} ++ (0,.47);
  \node[rectangle,
    draw,
    thin,
    minimum width = 7.2cm,
    minimum height = 5cm] (r) at (0,0) {};
  \node[text width=2cm, align=center, minimum size=0.5cm] (mic) at (-2.6,-0.3) {\scriptsize External\\[-0.6em] microphone};
  \node[text width=2cm, align=center, minimum size=0.5cm] (ls) at (-2.6,-1.8) {\scriptsize Loudspeaker};
  \node[text width=2cm, align=center, minimum size=0.5cm] (uav) at (1.63,1.9) {\scriptsize UAV};
\end{tikzpicture}
\caption{Side view}
\label{fig:drone:sketch1}
\end{subfigure}%
\hspace{.8cm}%
\begin{subfigure}{.4\textwidth}
  \centering
  \begin{tikzpicture}
    \pgfdeclareimage[width=.9\textwidth]{sketch_above}{./figures/DRONE_CONFERENCE_SKETCH_ABOVE1}
    \node (img2) at (0,0) {\pgfuseimage{sketch_above}};
    \node[rectangle,
      draw,
      thin,
      minimum width = 7.2cm,
      minimum height = 5cm] (r) at (0,0) {};
    \draw[|-|,line width=0.3mm] (-2.63,-2.1) -- node[below] {\scriptsize $2\,\text{m}$} ++ (4.26,0);
    \draw[dashed,line width=0.3mm] (-2.3,0) -- node[below] {} ++ (2.8,0);
    \node[text width=2cm, align=center, minimum size=0.5cm] (mic) at (-2.6,0.75) {\scriptsize External\\[-0.6em] microphone};
    \node[text width=2cm, align=center, minimum size=0.5cm] (ls) at (-2.6,-0.5) {\scriptsize Loudspeaker};
    \node[text width=2cm, align=center, minimum size=0.5cm] (uav) at (1.63,1.5) {\scriptsize UAV};
      \end{tikzpicture}
  \caption{Top view}
  \label{fig:drone:sketch:above1}
\end{subfigure}
\caption{Sketches of experimental setup}
\label{fig:setup:sketch1}
\end{figure}

\begin{figure}[h!]
  \centering
  \begin{tikzpicture}
    \node[anchor=south west,inner sep=0] at (0,0) {\includegraphics[trim=150 0 0 200,clip,angle=-90,width=.78\textwidth]{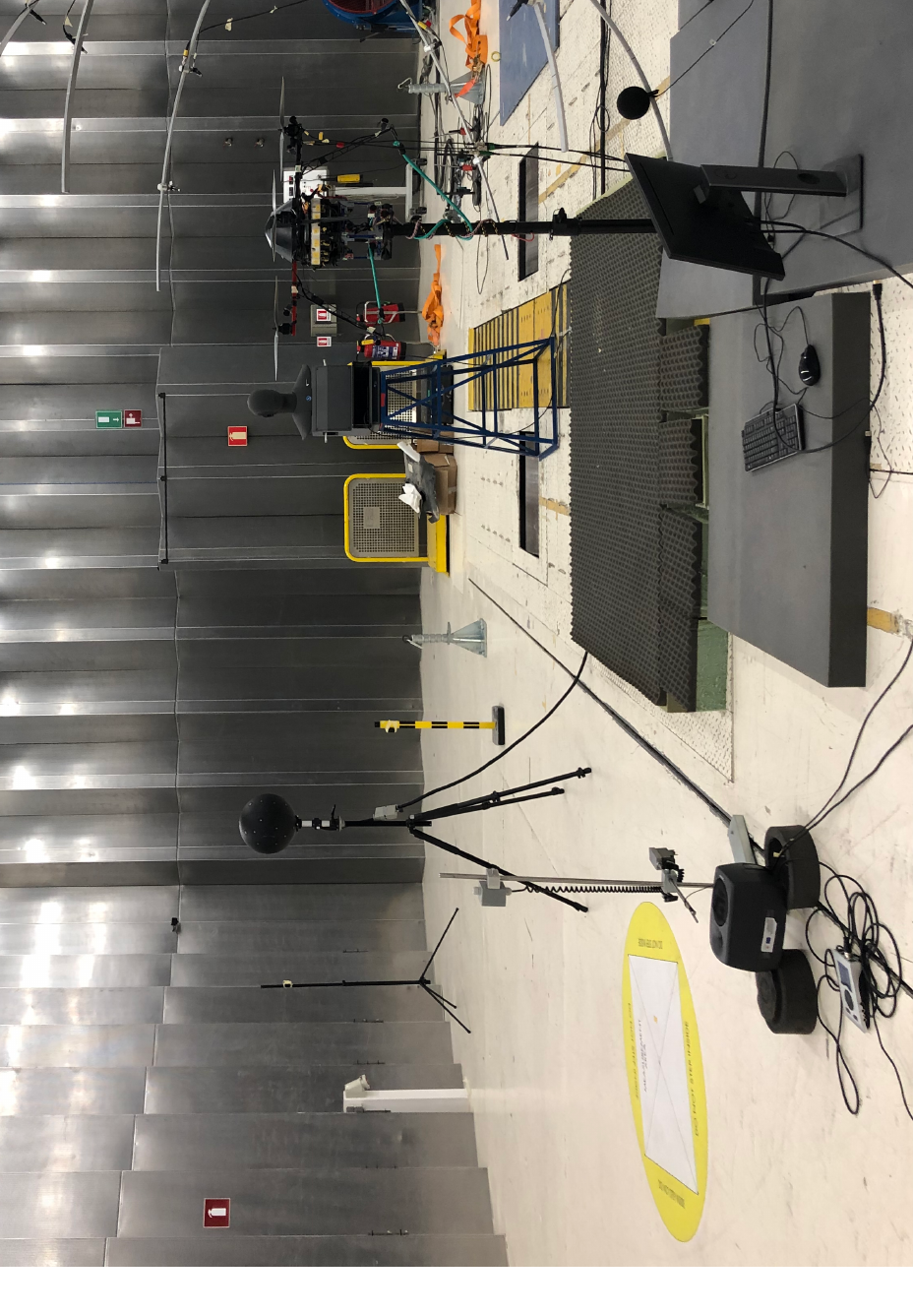}};
    \draw[green,ultra thick,rounded corners] (8.85,4.9) rectangle ++(3.45,1.55);
    \node[text width=2cm, align=center, minimum size=0.5cm, color=green] (uav) at (10.55,6.75) {\small UAV};
    \draw[magenta,ultra thick,rounded corners] (3.4,2.05) rectangle ++(0.9,0.4);
    \node[text width=2cm, align=center, minimum size=0.5cm, color=magenta] (mic) at (2.35,2.3) {\small External\\[-0.6em] microphone};
    \draw[blue,ultra thick,rounded corners] (3.1,1.1) rectangle ++(1.5,.9);
    \node[text width=2cm, align=center, minimum size=0.5cm, color=blue] (ls) at (1.95,1.65) {\small Loudspeaker};
  \end{tikzpicture}
  \caption{Experimental setup in semi-anechoic chamber}
  \label{fig:setup}
\end{figure}
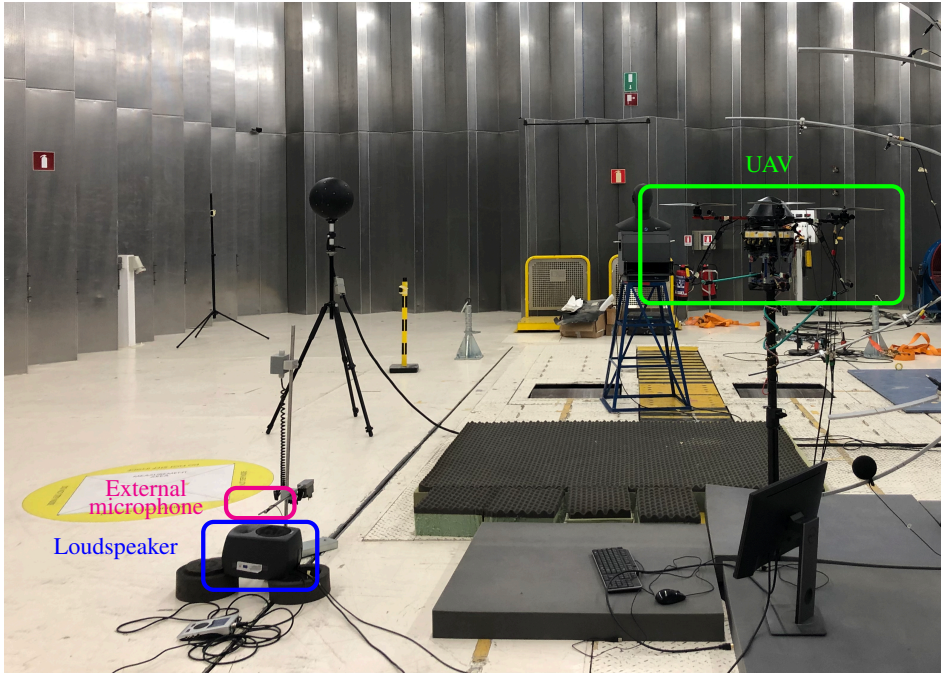

In the measurement setup considered for performance evaluation, a MikroKopter MK EASY Quadro V3 (HiSystems GmbH) quad-rotor UAV is mounted on a support stand such that the bottom of its custom-made frame is $1.15\,\text{m}$ above the floor. The UAV is equipped with a 16-element array composed of electret condenser microphones, a sound card and a minicomputer for performing audio recordings. A loudspeaker is placed $2\,\text{m}$ away from the base of the stand and close to ground level, in order to simulate a scenario of a speech source being present below the UAV's line-of-sight. An external microphone is placed $0.2\,\text{m}$ above the loudspeaker for recording the reference speech signal and allowing a performance comparison in the processing stage. A sketch representation is depicted in Fig.\,\ref{fig:setup:sketch1}, and a photo of the actual measurement setup inside the semi-anechoic chamber is presented in Fig.\,\ref{fig:setup}.

With the UAV's throttle level fixed to a nearly constant value, which was possible due to the remote control's throttle joystick not being spring-loaded, the setup conditions could be considered an approximation of a hovering state for the UAV \cite{hover}. A male speech signal from the VCTK corpus \cite{corpus} was played from the loudspeaker and recorded both by the external reference microphone and the drone's microphone array. The recordings, available online \cite{github}, were repeated for three different levels of throttle, and therefore, three different propeller rotational speeds, measured at run time with a laser probe placed under one of the propeller blades.

\subsection{Processing}
\label{sec:processing}
The measurements were processed offline on Matlab by, firstly, downsampling the recorded signals from $\SI{44.1}{\kilo\hertz}$ to $\SI{16}{\kilo\hertz}$. Then, a 512-point square-root of Hann observation window with 50\% overlap is employed for computing the FFT for the signal frames from each microphone. The correlation matrices $\hat{\mathbf{R}}_{\mathbf{yy}}$ and $\hat{\mathbf{R}}_{\mathbf{nn}}$ are estimated for each frequency bin based on the speech presence probability (SPP) \cite{spp} value of all frames being processed. Such SPP can be either estimated from one of the channels from the embedded microphone array (denoted as iSPP) or alternatively from the reference external microphone (denoted as xSPP), in order to provide a performance benchmark in the succeeding analysis of results obtained. We define $\beta(\kappa,l)$ as a speech activity indicator, which is computed as
\begin{align}
  \beta(\kappa,l)=
  \begin{cases}
      1 \text{ (speech active)}, &\text{if } \text{SPP}(\kappa,l)\geq 0.5\\
      0 \text{ (speech inactive)},&\text{otherwise}.
  \end{cases}
\end{align}
This parameter is then used to estimate the desired correlation matrices as
\begin{align}
  \hat{\mathbf{R}}_{\mathbf{yy}}(\kappa) &= \frac{1}{L_{\text{ON}}(\kappa)}\sum_{l=1}^{L}\mathbf{y}(\kappa,l)\mathbf{y}^{\mathrm{H}}(\kappa,l)\beta(\kappa,l),\\
    \hat{\mathbf{R}}_{\mathbf{nn}}(\kappa) &= \frac{1}{L_{\text{OFF}}(\kappa)}\sum_{l=1}^{L}\mathbf{y}(\kappa,l)\mathbf{y}^{\mathrm{H}}(\kappa,l)(1-\beta(\kappa,l)),
\end{align}
where $L$ denotes the total number of frames processed, and $L_{\text{ON}}(\kappa)$ and $L_{\text{OFF}}(\kappa)$ correspond to the total number of frames for frequency bin $\kappa$ where, based on the value of $\beta(\kappa,l)$ being 1 or 0, the desired speech component is assumed to be active or inactive, respectively.

In order to observe possible variations in performance according to different microphone array configurations, the MWF and PK-MWF filters are implemented considering three different numbers of channels used from the embedded microphone array, as illustrated in Fig.\,\ref{fig:coords}. In the PK-MWF implementation, the number of microphones from the main array used, denoted by $M_{\text{S+N}}$, corresponds to the number of channels employed in the selection matrix $\mathbf{H}$, whereas the four propeller microphones, placed below each pair of blades and denoted by $M_{\text{N}}$, are always used as noise references in the blocking matrix $\mathbf{B}$. In the case of MWF, however, we consider two implementations where the propeller microphones $M_{\text{N}}$ are used in its formulation or not, in addition to the channels used from the main array $M_{\text{S+N}}$.

The performance of the different methods implemented is evaluated in terms of SNR improvement, as well as of two other objective measures, namely the short-time objective intelligibility (STOI) \cite{stoi} and the perceptual evaluation of speech quality (PESQ) \cite{pesq}.

\definecolor{colorMp}{RGB}{0, 196, 0}%
\definecolor{colorMa_used}{RGB}{255,165,0}%
\definecolor{colorMa_unused}{rgb}{0.9,0.9,0.9}%
\begin{figure}[h!]
  \centering
  \begin{tikzpicture}
\begin{customlegend}[legend columns=3,legend style={align=left,column sep=3ex},
      legend entries={\hspace{-1em}\small Propeller microphones ($M_{\text{N}}$),
                      \hspace{-1em}\small Used main array microphones ($M_{\text{S+N}}$),
                      \hspace{-1em}\small Unused main array microphones
                      }]
      \addlegendimage{line width=0.5pt, mark=*, mark options={fill=colorMp},only marks}
      \addlegendimage{line width=0.5pt, mark=*, mark options={fill=colorMa_used}, only marks}
      \addlegendimage{line width=0.5pt, mark=*, mark options={fill=colorMa_unused}, only marks}
      \end{customlegend}
\end{tikzpicture}\\
\vspace{0.2cm}
  \begin{subfigure}{.33\columnwidth}
    \centering
    \includegraphics[width=\columnwidth]{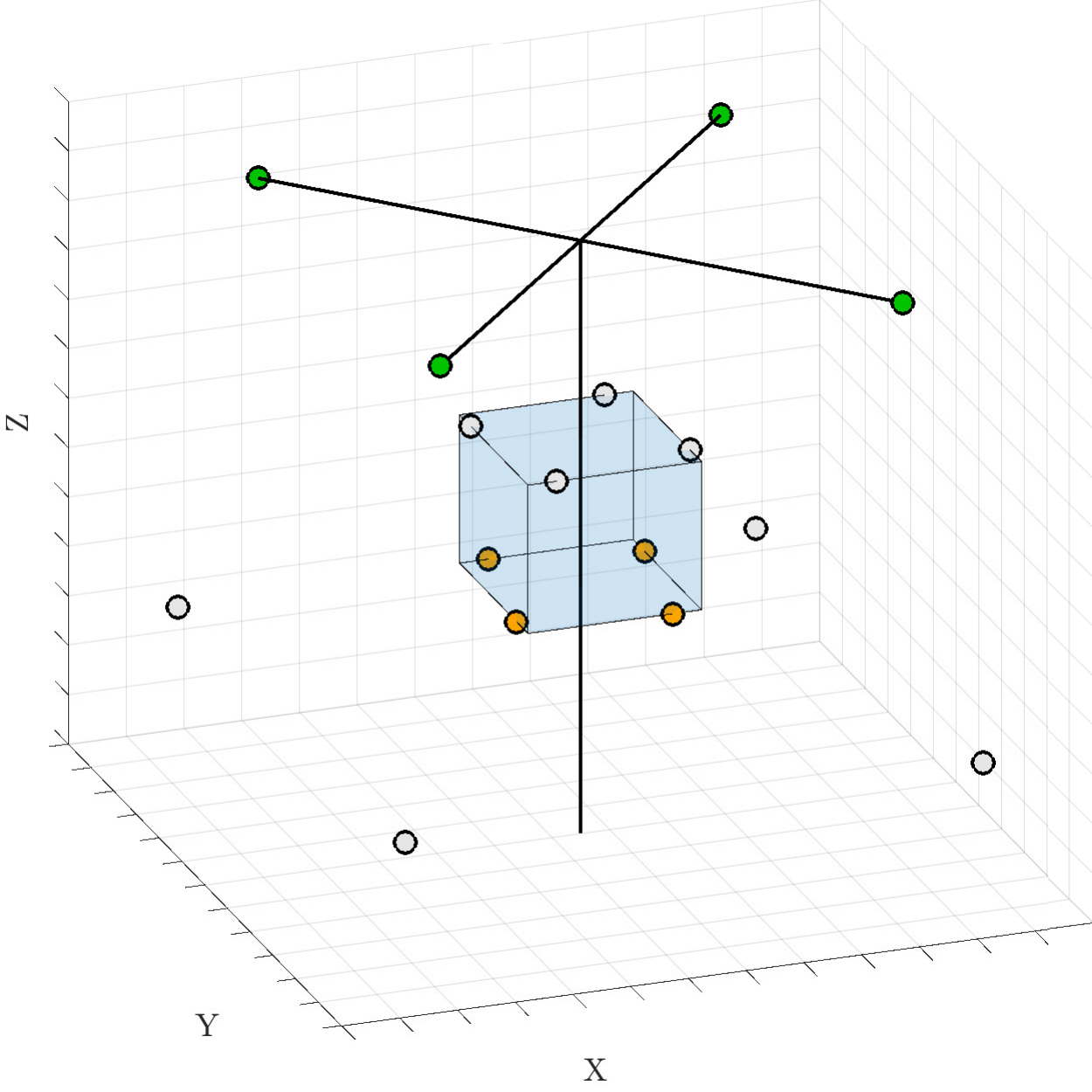}
    \caption{$M_{\text{S+N}}=4$, $M_{\text{N}}=4$}
    \label{fig:coords:8ch}
  \end{subfigure}%
  \begin{subfigure}{.33\columnwidth}
    \centering
    \includegraphics[width=\columnwidth]{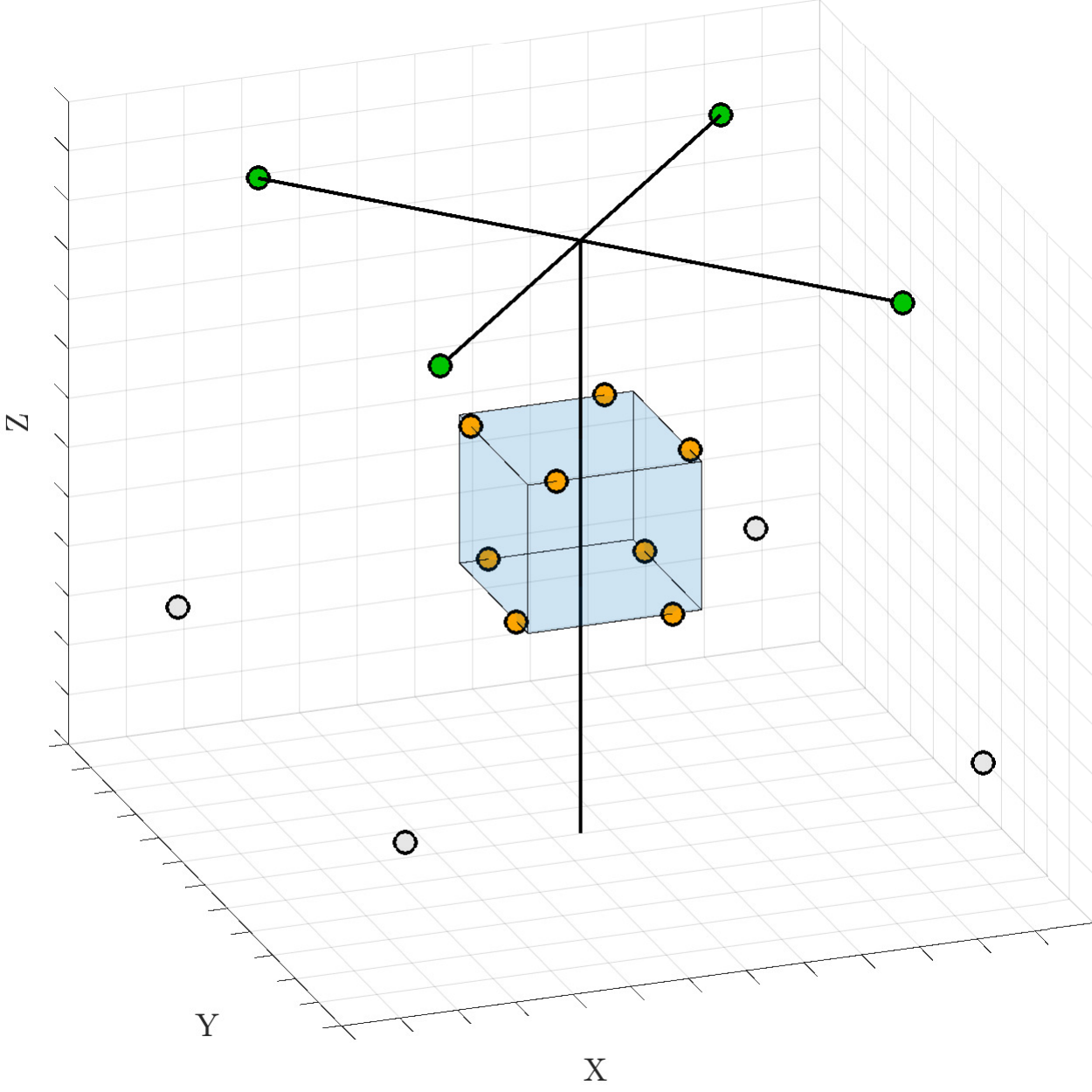}
    \caption{$M_{\text{S+N}}=8$, $M_{\text{N}}=4$}
    \label{fig:coords:12ch}
  \end{subfigure}%
  \begin{subfigure}{.33\columnwidth}
    \centering
    \includegraphics[width=\columnwidth]{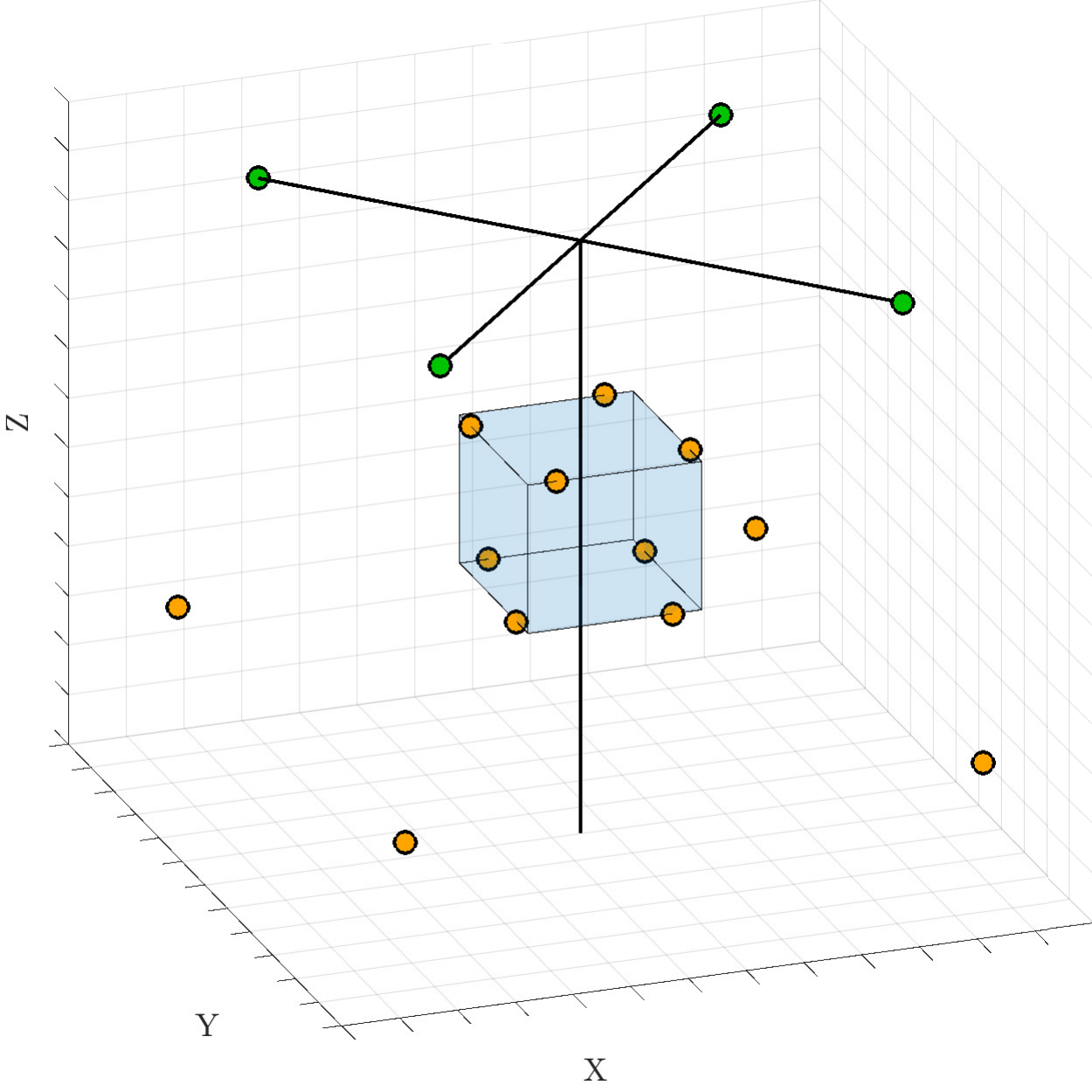}
    \caption{$M_{\text{S+N}}=12$, $M_{\text{N}}=4$}
    \label{fig:coords:16ch}
  \end{subfigure}%
  \caption{Illustration of different microphone array configurations considered in the formulation of MWF and PK-MWF.}
  \label{fig:coords}
\end{figure}

\section{\uppercase{Results}}
\label{sec:results}
The resulting SNR improvement obtained at the channel where the reference signal estimate $\hat{d}=\hat{\mathbf{w}}^{\mathrm{H}}\mathbf{y}$ is computed is presented in Fig.\,\ref{fig:snr}. Different numbers of microphones in the proposed filtering formulations and the SPP estimates were considered. The bottom and top horizontal axes indicate the original SNRs considered and the corresponding average rotational speed of the rotors in revolutions/min (rpm), respectively. We can observe that, with the SPP estimate from the reference external signal (xSPP), the SNR improvement is greater than when the SPP estimate from the embedded array's own reference channel (iSPP) is used, as the external microphone placed near the target source provides a signal with greater SNR itself. Therefore, the detection of speech activity is more reliable, resulting in more accurate estimates of the necessary correlations matrices here considered. Since having an external microphone near the target speaker may not be realistic in all kinds of scenarios, the use of the external SPP estimate can be here seen as a way to provide a performance benchmark for the proposed methods, with which it is possible to compare and evaluate the improvements obtained and its possible limitations.
In this case, it is observed that even with a poorer SPP estimate, all methods are still capable of improving the SNR of the target reference channel.

In terms of the number of microphones used for performing the multichannel filtering, it is observed that performance is improved when using more microphone signals, although the differences are more visible when going from $M_{\text{S+N}} = 4$ to $M_{\text{S+N}}=8$ than when going from $M_{\text{S+N}}=8$ to $M_{\text{S+N}}=12$. This is possibly due to performance saturation related to signal model or SPP errors that could not be improved with the increase in number of microphones used, as well as to the microphone placement of the last four channels included being aligned with the rotors' rotational axes, which is considered to be where the noise level is the strongest \cite{book_hioka} and therefore, not providing as much additional information on the target speech signal as the previously included microphones.

We can also observe that the performance of PK-MWF in terms of SNR improvement is similar to the one of MWF while employing the same number of microphones ($M=M_{\text{S+N}}+M_{\text{N}}$). However, the matrix reductions performed in the PK-MWF might favor its usage in terms of computational complexity.

The improvement in speech intelligibility measured by STOI is presented in Fig.\,\ref{fig:stoi}. It can be observed that the performance of PK-MWF with respect to this perceptual measure is particularly better than both formulations of standard MWF considered when employing the SPP estimate from the embedded array's own reference signal (iSPP), indicating the advantage of using the ego-noise reference signals as proposed here for improving robustness to SPP errors which affect the estimation of $\hat{\mathbf{R}}_{\mathbf{yy}}$ and $\hat{\mathbf{R}}_{\mathbf{nn}}$. In addition to having a similar behavior in performance improvement when including more channels as the one seen in terms of SNR, we can also observe that the STOI improvement rate increases with the original values from the array's reference signal, which corresponds to a decrease in the average rotor speed.

\definecolor{mycolor1}{rgb}{0.00000,0.44700,0.74100}%
\definecolor{mycolor2}{rgb}{0.85000,0.32500,0.09800}%
\definecolor{mycolor3}{rgb}{0.92900,0.69400,0.12500}%
\definecolor{mycolor4}{rgb}{0.00000,0.44706,0.74118}%
\definecolor{mycolor5}{rgb}{0.85098,0.32549,0.09804}%
\definecolor{mycolor6}{rgb}{0.92941,0.69412,0.12549}%
\begin{figure}[ht!]
  \centering
  \begin{tikzpicture}
\begin{customlegend}[legend columns=3,legend style={align=left,column sep=3ex},
      legend entries={\hspace{-1em}\small MWF ($M=M_{\text{S+N}}$) - iSPP,
                      \hspace{-1em}\small MWF ($M=M_{\text{S+N}}+M_{\text{N}}$) - iSPP,
                      \hspace{-1em}\small PK-MWF ($M=M_{\text{S+N}}+M_{\text{N}}$) - iSPP,
                      \hspace{-1em}\small MWF ($M=M_{\text{S+N}}$) - xSPP,
                      \hspace{-1em}\small MWF ($M=M_{\text{S+N}}+M_{\text{N}}$) - xSPP,
                      \hspace{-1em}\small PK-MWF ($M=M_{\text{S+N}}+M_{\text{N}}$) - xSPP
                      }]
      \addlegendimage{color=mycolor1, dashed, line width=2.0pt, mark=o, mark options={solid, mycolor1}}
      \addlegendimage{color=mycolor2, dashed, line width=2.0pt, mark=o, mark options={solid, mycolor2}}
      \addlegendimage{color=mycolor3, dashed, line width=2.0pt, mark=o, mark options={solid, mycolor3}}
      \addlegendimage{color=mycolor4, line width=2.0pt, mark=o, mark options={solid, mycolor4}}
      \addlegendimage{color=mycolor5, line width=2.0pt, mark=o, mark options={solid, mycolor5}}
      \addlegendimage{color=mycolor6, line width=2.0pt, mark=o, mark options={solid, mycolor6}}
      \end{customlegend}
\end{tikzpicture}\\
\vspace{0.2cm}
  \newlength\fheight
\newlength\fwidth
\setlength\fheight{.30\columnwidth}
\setlength\fwidth{.22\columnwidth}
    \begin{subfigure}{.30\columnwidth}
        \centering
        \input{./plots/snr_Ma_4.tex}
        \caption{$M_{\text{S+N}} = 4$}
        \label{fig:snr:4}
    \end{subfigure}
    \hfill%
    \begin{subfigure}{.30\columnwidth}
        \centering
        \input{./plots/snr_Ma_8.tex}
        \caption{$M_{\text{S+N}} = 8$}
        \label{fig:snr:8}
    \end{subfigure}
    \hfill%
    \begin{subfigure}{.30\columnwidth}
        \centering
        \input{./plots/snr_Ma_12.tex}
        \caption{$M_{\text{S+N}} = 12$}
        \label{fig:snr:12}
    \end{subfigure}
    \caption{SNR improvement in function of original SNRs associated to different average rotor speeds.}
    \label{fig:snr}
\end{figure}
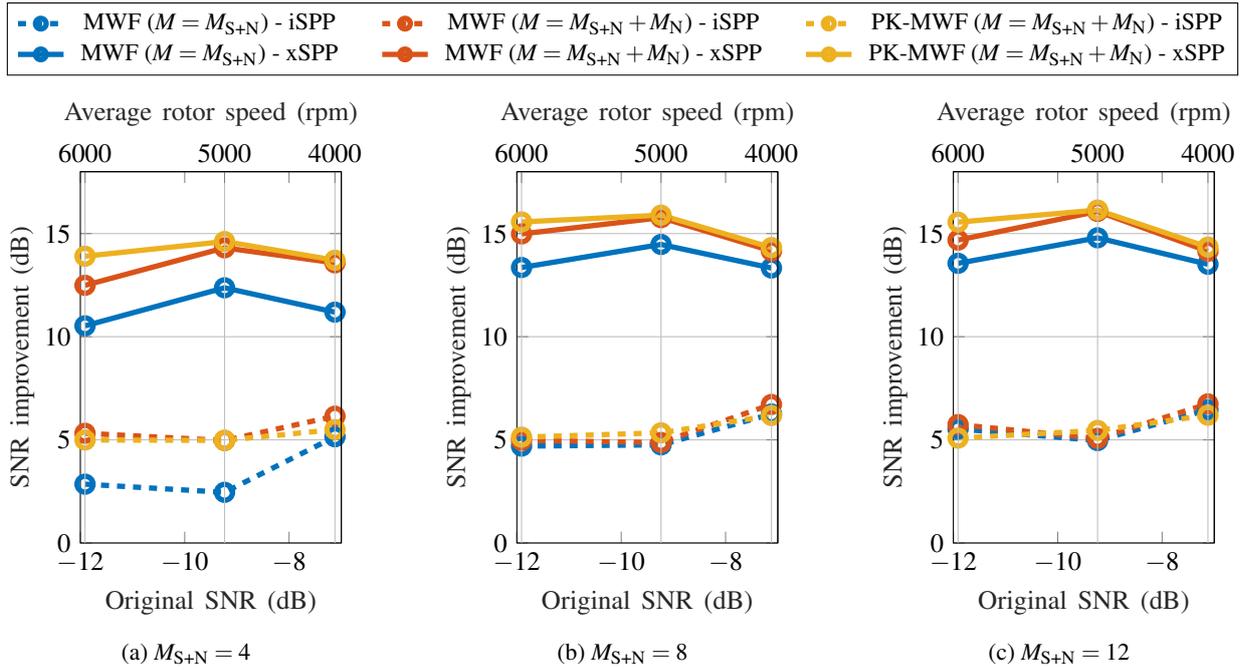

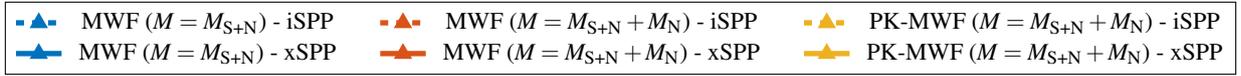
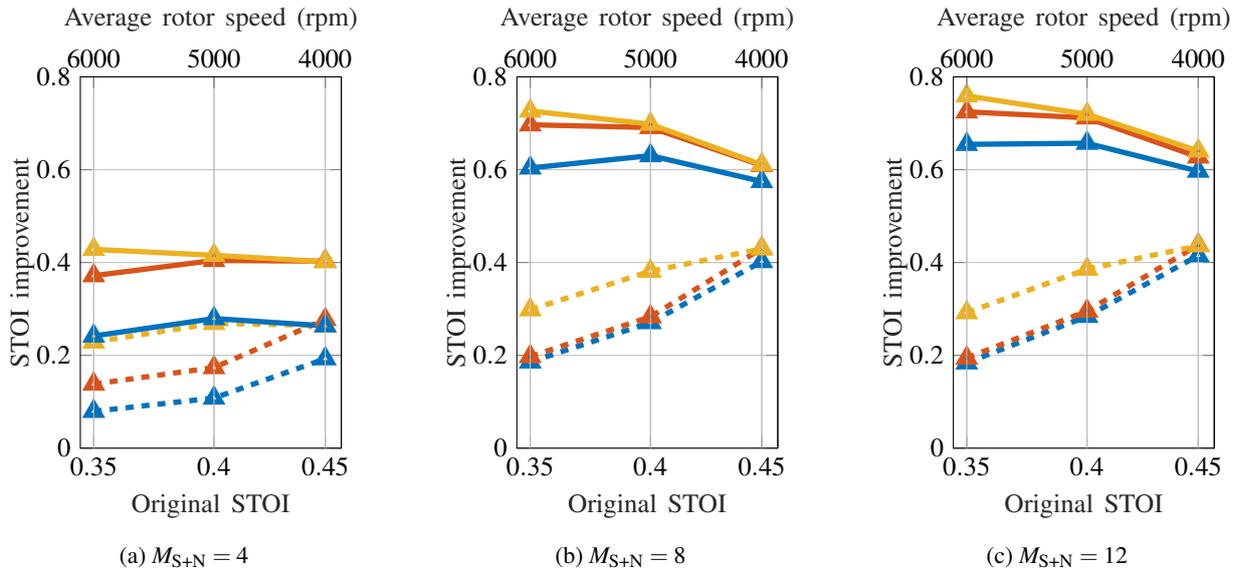
\begin{figure}[b!]
  \centering
  \begin{tikzpicture}
\begin{customlegend}[legend columns=3,legend style={align=left,column sep=3ex},
      legend entries={\hspace{-1em}\small MWF ($M=M_{\text{S+N}}$) - iSPP,
                      \hspace{-1em}\small MWF ($M=M_{\text{S+N}}+M_{\text{N}}$) - iSPP,
                      \hspace{-1em}\small PK-MWF ($M=M_{\text{S+N}}+M_{\text{N}}$) - iSPP,
                      \hspace{-1em}\small MWF ($M=M_{\text{S+N}}$) - xSPP,
                      \hspace{-1em}\small MWF ($M=M_{\text{S+N}}+M_{\text{N}}$) - xSPP,
                      \hspace{-1em}\small PK-MWF ($M=M_{\text{S+N}}+M_{\text{N}}$) - xSPP
                      }]
      \addlegendimage{color=mycolor1, dashed, line width=2.0pt, mark=triangle, mark options={solid, mycolor1}}
      \addlegendimage{color=mycolor2, dashed, line width=2.0pt, mark=triangle, mark options={solid, mycolor2}}
      \addlegendimage{color=mycolor3, dashed, line width=2.0pt, mark=triangle, mark options={solid, mycolor3}}
      \addlegendimage{color=mycolor4, line width=2.0pt, mark=triangle, mark options={solid, mycolor4}}
      \addlegendimage{color=mycolor5, line width=2.0pt, mark=triangle, mark options={solid, mycolor5}}
      \addlegendimage{color=mycolor6, line width=2.0pt, mark=triangle, mark options={solid, mycolor6}}
      \end{customlegend}
\end{tikzpicture}\\
\vspace{0.2cm}
\setlength\fheight{.30\columnwidth}
\setlength\fwidth{.22\columnwidth}
    \begin{subfigure}{.30\columnwidth}
        \centering
        \input{./plots/stoi_improv_Ma_4.tex}
        \caption{$M_{\text{S+N}} = 4$}
        \label{fig:stoi2:4}
    \end{subfigure}
    \hfill%
    \begin{subfigure}{.30\columnwidth}
        \centering
        \input{./plots/stoi_improv_Ma_8.tex}
        \caption{$M_{\text{S+N}} = 8$}
        \label{fig:stoi2:8}
    \end{subfigure}
    \hfill%
    \begin{subfigure}{.30\columnwidth}
        \centering
        \input{./plots/stoi_improv_Ma_12.tex}
        \caption{$M_{\text{S+N}} = 12$}
        \label{fig:stoi2:12}
    \end{subfigure}
    \caption{STOI improvement in function of original STOI values associated to different average rotor speeds.}
    \label{fig:stoi}
\end{figure}

\begin{figure}[b!]
  \centering
  \begin{tikzpicture}
\begin{customlegend}[legend columns=3,legend style={align=left,column sep=3ex},
      legend entries={\hspace{-1em}\small MWF ($M=M_{\text{S+N}}$) - iSPP,
                      \hspace{-1em}\small MWF ($M=M_{\text{S+N}}+M_{\text{N}}$) - iSPP,
                      \hspace{-1em}\small PK-MWF ($M=M_{\text{S+N}}+M_{\text{N}}$) - iSPP,
                      \hspace{-1em}\small MWF ($M=M_{\text{S+N}}$) - xSPP,
                      \hspace{-1em}\small MWF ($M=M_{\text{S+N}}+M_{\text{N}}$) - xSPP,
                      \hspace{-1em}\small PK-MWF ($M=M_{\text{S+N}}+M_{\text{N}}$) - xSPP
                      }]
      \addlegendimage{color=mycolor1, dashed, line width=2.0pt, mark=square, mark options={solid, mycolor1}}
      \addlegendimage{color=mycolor2, dashed, line width=2.0pt, mark=square, mark options={solid, mycolor2}}
      \addlegendimage{color=mycolor3, dashed, line width=2.0pt, mark=square, mark options={solid, mycolor3}}
      \addlegendimage{color=mycolor4, line width=2.0pt, mark=square, mark options={solid, mycolor4}}
      \addlegendimage{color=mycolor5, line width=2.0pt, mark=square, mark options={solid, mycolor5}}
      \addlegendimage{color=mycolor6, line width=2.0pt, mark=square, mark options={solid, mycolor6}}
      \end{customlegend}
\end{tikzpicture}\\
\vspace{0.2cm}
\setlength\fheight{.30\columnwidth}
\setlength\fwidth{.22\columnwidth}
    \begin{subfigure}{.30\columnwidth}
        \centering
        \input{./plots/pesq_improv_Ma_4.tex}
        \caption{$M_{\text{S+N}} = 4$}
        \label{fig:pesq1:4}
    \end{subfigure}
    \hfill%
    \begin{subfigure}{.30\columnwidth}
        \centering
        \input{./plots/pesq_improv_Ma_8.tex}
        \caption{$M_{\text{S+N}} = 8$}
        \label{fig:pesq1:8}
    \end{subfigure}
    \hfill%
    \begin{subfigure}{.30\columnwidth}
        \centering
        \input{./plots/pesq_improv_Ma_12.tex}
        \caption{$M_{\text{S+N}} = 12$}
        \label{fig:pesq1:12}
    \end{subfigure}
    \caption{PESQ improvement in function of original PESQ scores associated with different average rotor speeds.}
    \label{fig:pesq}
\end{figure}
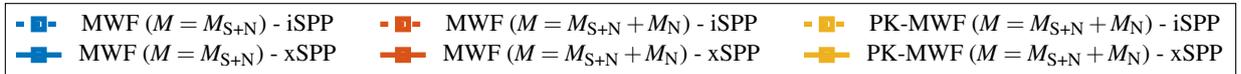
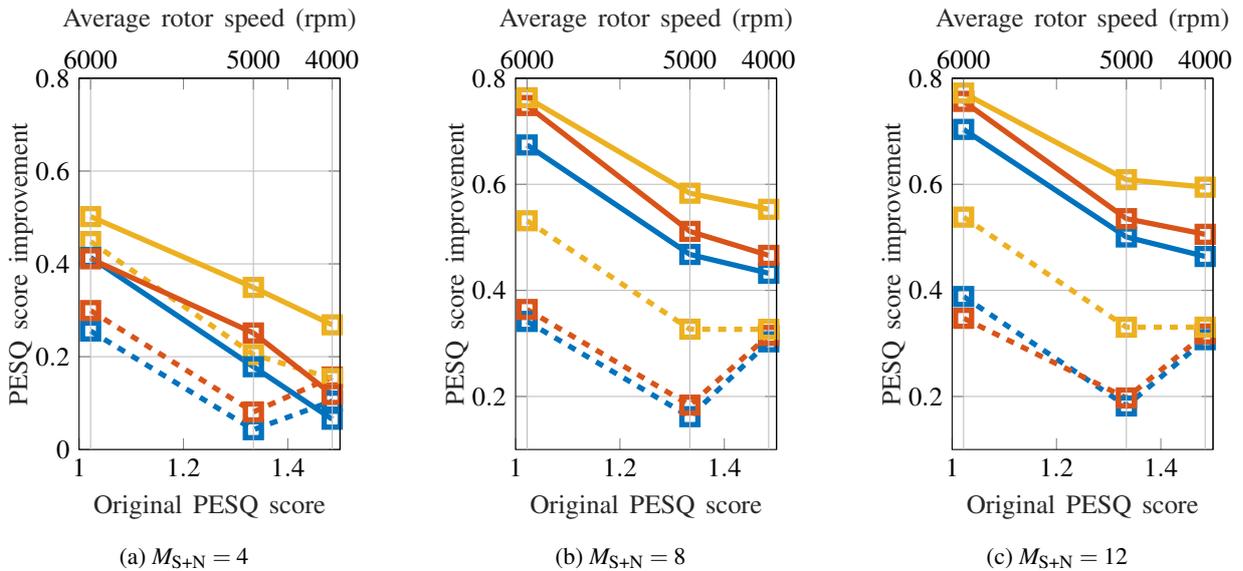

Finally, the improvement in speech quality measured by PESQ is presented in Fig.\,\ref{fig:pesq}, which indicates a better performance of PK-MWF compared to the standard MWF filters when using either one of the SPP estimates considered. As opposed to the STOI improvement depicted in Fig.\,\ref{fig:stoi}, the performance improvement rate in this case decreases with the original PESQ score of the reference signal considered, which is in turn associated to an increase in the average rotational speed of the rotors.

\section{\uppercase{Conclusion}}
\label{sec:conclusion}
In this paper, a method for speech enhancement using ego-noise references with a microphone array embedded in an unmanned aerial vehicle was proposed. The ego-noise reference signals captured by microphones placed near the UAV's rotors were used as prior knowledge in the PK-MWF for estimating the necessary speech correlation matrix in a constrained optimization problem. Speech presence probability (SPP) was estimated from both an external reference microphone and one of the embedded array's channels in order to detect speech activity. Results obtained from experimental recordings showed that, especially in a more realistic scenario where the SPP is estimated from one of the UAV's own microphones, the PK-MWF implementation provided greater improvement in speech intelligibility and quality when compared to the use of standard MWF, with and without the propeller microphones being included in its formulation. While the SNR improvement is similar for PK-MWF and the standard MWF using the propeller microphones, it can be argued that employing the PK-MWF implementation might be advantageous in terms of computational complexity due to the performed matrix reductions when considering real-time processing applications. Future work includes expanding the measurement campaign to consider moving UAVs, updating the correlation matrices over different time frames, and reformulating the noise signal model to consider more particular characteristics of UAVs' ego-noise.

\section*{\uppercase{Acknowledgements}}

This research work was carried out at the ESAT Laboratory of KU Leuven, in the frame of FWO Mandate: SB 1S86520N. The research leading to these results has received funding from the European Research Council under the European Union's Horizon 2020 research and innovation program / ERC Consolidator Grant: SONORA (no. 773268), the H2020 MSCA ITN ETN PBNv2 project (GA 721615) and the H2020 MSCA ITN ETN VRACE project (GA 812719). This paper reflects only the authors' views and the Union is not liable for any use that may be made of the contained information. The authors also wish to thank Mr. Sacha Morales for his support in the measurement campaign.


\bibliographystyle{abbrv}
\renewcommand{\refname}{\normalfont\selectfont\normalsize}

\end{document}

%% file: plots/snr_Ma_4.tex
%
%
\definecolor{mycolor1}{rgb}{0.00000,0.44700,0.74100}%
\definecolor{mycolor2}{rgb}{0.85000,0.32500,0.09800}%
\definecolor{mycolor3}{rgb}{0.92900,0.69400,0.12500}%
\definecolor{mycolor4}{rgb}{0.00000,0.44706,0.74118}%
\definecolor{mycolor5}{rgb}{0.85098,0.32549,0.09804}%
\definecolor{mycolor6}{rgb}{0.92941,0.69412,0.12549}%
\begin{tikzpicture}

\begin{axis}[%
width=0.951\fwidth,
height=\fheight,
at={(0\fwidth,0\fheight)},
scale only axis,
xmin=-12,
xmax=-7,
xlabel style={font=\color{white!15!black}},
xlabel={Original SNR (dB)},
ymin=0,
ymax=18,
ylabel style={at={(axis description cs:-.14,.5)},anchor=south,font=\color{white!15!black}},
ylabel={SNR improvement (dB)},
axis background/.style={fill=white},
ymajorgrids,
legend style={legend cell align=left, align=left, draw=white!15!black}
]

\addplot [color=mycolor1, dashed, line width=2.0pt, mark size=3.0pt, mark=o, mark options={solid, mycolor1}]
  table[row sep=crcr]{%
-11.9101246344674	2.85106344555103\\
-9.23665478296534	2.45687640118589\\
-7.12156128075881	5.14877925443668\\
};

\addplot [color=mycolor2, dashed, line width=2.0pt, mark size=3.0pt, mark=o, mark options={solid, mycolor2}]
  table[row sep=crcr]{%
-11.9101246344674	5.31814426896753\\
-9.23665478296534	4.96871799229644\\
-7.12156128075881	6.1463479034233\\
};

\addplot [color=mycolor3, dashed, line width=2.0pt, mark size=3.0pt, mark=o, mark options={solid, mycolor3}]
  table[row sep=crcr]{%
-11.9101246344674	4.99411877657407\\
-9.23665478296534	4.97498412193391\\
-7.12156128075881	5.49653771893025\\
};

\addplot [color=mycolor4, line width=2.0pt, mark size=3.0pt, mark=o, mark options={solid, mycolor4}]
  table[row sep=crcr]{%
-11.9101246344674	10.5278673532475\\
-9.23665478296534	12.3719169472136\\
-7.12156128075881	11.1909792414659\\
};

\addplot [color=mycolor5, line width=2.0pt, mark size=3.0pt, mark=o, mark options={solid, mycolor5}]
  table[row sep=crcr]{%
-11.9101246344674	12.4917561507856\\
-9.23665478296534	14.3091732983461\\
-7.12156128075881	13.5801057804215\\
};

\addplot [color=mycolor6, line width=2.0pt, mark size=3.0pt, mark=o, mark options={solid, mycolor6}]
  table[row sep=crcr]{%
-11.9101246344674	13.9017717865176\\
-9.23665478296534	14.6141559494895\\
-7.12156128075881	13.7099184338271\\
};

\end{axis}

\begin{axis}[%
width=0.951\fwidth,
height=\fheight,
at={(0\fwidth,0\fheight)},
scale only axis,
xmin=-12,
xmax=-7,
xtick={-11.9101246344674,-9.23665478296534,-7.12156128075881},
xticklabels={{6000},{5000},{4000}},
xlabel style={font=\color{white!15!black}},
xlabel={Average rotor speed (rpm)},
ymin=0,
ymax=1,
ytick={\empty},
axis x line*=top,
axis y line*=right,
xmajorgrids,
ymajorgrids,
legend style={legend cell align=left, align=left, draw=white!15!black, fill=white!94!black}
]
\end{axis}
\end{tikzpicture}%

%% file: plots/snr_Ma_8.tex
%
%
\definecolor{mycolor1}{rgb}{0.00000,0.44700,0.74100}%
\definecolor{mycolor2}{rgb}{0.85000,0.32500,0.09800}%
\definecolor{mycolor3}{rgb}{0.92900,0.69400,0.12500}%
\definecolor{mycolor4}{rgb}{0.00000,0.44706,0.74118}%
\definecolor{mycolor5}{rgb}{0.85098,0.32549,0.09804}%
\definecolor{mycolor6}{rgb}{0.92941,0.69412,0.12549}%
\begin{tikzpicture}

\begin{axis}[%
width=0.951\fwidth,
height=\fheight,
at={(0\fwidth,0\fheight)},
scale only axis,
xmin=-12,
xmax=-7,
xlabel style={font=\color{white!15!black}},
xlabel={Original SNR (dB)},
ymin=0,
ymax=18,
ylabel style={at={(axis description cs:-.14,.5)},anchor=south,font=\color{white!15!black}},
ylabel={SNR improvement (dB)},
axis background/.style={fill=white},
ymajorgrids,
legend style={legend cell align=left, align=left, draw=white!15!black}
]
\addplot [color=mycolor1, dashed, line width=2.0pt, mark size=3.0pt, mark=o, mark options={solid, mycolor1}]
  table[row sep=crcr]{%
-11.9101246344674	4.69625291311993\\
-9.23665478296534	4.75637396232521\\
-7.12156128075881	6.25366961804684\\
};

\addplot [color=mycolor2, dashed, line width=2.0pt, mark size=3.0pt, mark=o, mark options={solid, mycolor2}]
  table[row sep=crcr]{%
-11.9101246344674	4.97659432296729\\
-9.23665478296534	4.87219038153233\\
-7.12156128075881	6.71639696897687\\
};

\addplot [color=mycolor3, dashed, line width=2.0pt, mark size=3.0pt, mark=o, mark options={solid, mycolor3}]
  table[row sep=crcr]{%
-11.9101246344674	5.13008695091014\\
-9.23665478296534	5.3405780070948\\
-7.12156128075881	6.18055590779131\\
};

\addplot [color=mycolor4, line width=2.0pt, mark size=3.0pt, mark=o, mark options={solid, mycolor4}]
  table[row sep=crcr]{%
-11.9101246344674	13.3552717720631\\
-9.23665478296534	14.4726178422706\\
-7.12156128075881	13.3316792839585\\
};

\addplot [color=mycolor5, line width=2.0pt, mark size=3.0pt, mark=o, mark options={solid, mycolor5}]
  table[row sep=crcr]{%
-11.9101246344674	14.9836199604146\\
-9.23665478296534	15.7700776796636\\
-7.12156128075881	14.1573493601626\\
};

\addplot [color=mycolor6, line width=2.0pt, mark size=3.0pt, mark=o, mark options={solid, mycolor6}]
  table[row sep=crcr]{%
-11.9101246344674	15.5647154379141\\
-9.23665478296534	15.8939076669485\\
-7.12156128075881	14.3211865444645\\
};

\end{axis}

\begin{axis}[%
width=0.951\fwidth,
height=\fheight,
at={(0\fwidth,0\fheight)},
scale only axis,
xmin=-12,
xmax=-7,
xtick={-11.9101246344674,-9.23665478296534,-7.12156128075881},
xticklabels={{6000},{5000},{4000}},
xlabel style={font=\color{white!15!black}},
xlabel={Average rotor speed (rpm)},
ymin=0,
ymax=1,
ytick={\empty},
axis x line*=top,
axis y line*=right,
xmajorgrids,
ymajorgrids,
legend style={legend cell align=left, align=left, draw=white!15!black, fill=white!94!black}
]
ylabel style={at={(axis description cs:-.15,.5)},anchor=south,font=\color{white!15!black}},
\end{axis}
\end{tikzpicture}%

%% file: plots/snr_Ma_12.tex
%
%
\definecolor{mycolor1}{rgb}{0.00000,0.44700,0.74100}%
\definecolor{mycolor2}{rgb}{0.85000,0.32500,0.09800}%
\definecolor{mycolor3}{rgb}{0.92900,0.69400,0.12500}%
\definecolor{mycolor4}{rgb}{0.00000,0.44706,0.74118}%
\definecolor{mycolor5}{rgb}{0.85098,0.32549,0.09804}%
\definecolor{mycolor6}{rgb}{0.92941,0.69412,0.12549}%
\begin{tikzpicture}

\begin{axis}[%
width=0.951\fwidth,
height=\fheight,
at={(0\fwidth,0\fheight)},
scale only axis,
xmin=-12,
xmax=-7,
xlabel style={font=\color{white!15!black}},
xlabel={Original SNR (dB)},
ymin=0,
ymax=18,
ylabel style={at={(axis description cs:-.14,.5)},anchor=south,font=\color{white!15!black}},
ylabel={SNR improvement (dB)},
axis background/.style={fill=white},
ymajorgrids,
legend style={legend cell align=left, align=left, draw=white!15!black}
]

\addplot [color=mycolor1, dashed, line width=2.0pt, mark size=3.0pt, mark=o, mark options={solid, mycolor1}]
  table[row sep=crcr]{%
-11.9101246344674	5.53019144853681\\
-9.23665478296534	4.96618601830948\\
-7.12156128075881	6.45808376178055\\
};

\addplot [color=mycolor2, dashed, line width=2.0pt, mark size=3.0pt, mark=o, mark options={solid, mycolor2}]
  table[row sep=crcr]{%
-11.9101246344674	5.74027047002251\\
-9.23665478296534	5.07011925610335\\
-7.12156128075881	6.74335929737131\\
};

\addplot [color=mycolor3, dashed, line width=2.0pt, mark size=3.0pt, mark=o, mark options={solid, mycolor3}]
  table[row sep=crcr]{%
-11.9101246344674	5.08507680222384\\
-9.23665478296534	5.45793175169764\\
-7.12156128075881	6.19451863416011\\
};

\addplot [color=mycolor4, line width=2.0pt, mark size=3.0pt, mark=o, mark options={solid, mycolor4}]
  table[row sep=crcr]{%
-11.9101246344674	13.5584278461398\\
-9.23665478296534	14.7953644339911\\
-7.12156128075881	13.5119439976216\\
};

\addplot [color=mycolor5, line width=2.0pt, mark size=3.0pt, mark=o, mark options={solid, mycolor5}]
  table[row sep=crcr]{%
-11.9101246344674	14.6844378644356\\
-9.23665478296534	16.0748981356333\\
-7.12156128075881	14.1440511859926\\
};

\addplot [color=mycolor6, line width=2.0pt, mark size=3.0pt, mark=o, mark options={solid, mycolor6}]
  table[row sep=crcr]{%
-11.9101246344674	15.5547529222294\\
-9.23665478296534	16.1405330900547\\
-7.12156128075881	14.3451693655331\\
};

\end{axis}

\begin{axis}[%
width=0.951\fwidth,
height=\fheight,
at={(0\fwidth,0\fheight)},
scale only axis,
xmin=-12,
xmax=-7,
xtick={-11.9101246344674,-9.23665478296534,-7.12156128075881},
xticklabels={{6000},{5000},{4000}},
xlabel style={font=\color{white!15!black}},
xlabel={Average rotor speed (rpm)},
ymin=0,
ymax=1,
ytick={\empty},
axis x line*=top,
axis y line*=right,
xmajorgrids,
ymajorgrids,
legend style={legend cell align=left, align=left, draw=white!15!black, fill=white!94!black}
]
ylabel style={at={(axis description cs:-.15,.5)},anchor=south,font=\color{white!15!black}},
\end{axis}
\end{tikzpicture}%

%% file: plots/stoi_improv_Ma_4.tex
%
%
\definecolor{mycolor1}{rgb}{0.00000,0.44700,0.74100}%
\definecolor{mycolor2}{rgb}{0.85000,0.32500,0.09800}%
\definecolor{mycolor3}{rgb}{0.92900,0.69400,0.12500}%
\definecolor{mycolor4}{rgb}{0.00000,0.44706,0.74118}%
\definecolor{mycolor5}{rgb}{0.85098,0.32549,0.09804}%
\definecolor{mycolor6}{rgb}{0.92941,0.69412,0.12549}%
\begin{tikzpicture}

\begin{axis}[%
width=0.951\fwidth,
height=\fheight,
at={(0\fwidth,0\fheight)},
scale only axis,
xmin=0.34,
xmax=0.455,
xlabel style={font=\color{white!15!black}},
xlabel={Original STOI},
ymin=0,
ymax=0.8,
ylabel style={at={(axis description cs:-.14,.5)},anchor=south,font=\color{white!15!black}},
ylabel={STOI improvement},
axis background/.style={fill=white},
xmajorgrids,
ymajorgrids,
xtick={0.345895072941106,0.398914485462443,0.447944293311666},
]
\addplot [color=mycolor1, dashed, line width=2.0pt, mark size=3.0pt, mark=triangle, mark options={solid, mycolor1}]
  table[row sep=crcr]{%
0.345895072941106	0.0793655733263559\\
0.398914485462443	0.108007243226739\\
0.447944293311666	0.192711387280928\\
};

\addplot [color=mycolor2, dashed, line width=2.0pt, mark size=3.0pt, mark=triangle, mark options={solid, mycolor2}]
  table[row sep=crcr]{%
0.345895072941106	0.138121036110393\\
0.398914485462443	0.173643972685684\\
0.447944293311666	0.277258128152999\\
};

\addplot [color=mycolor3, dashed, line width=2.0pt, mark size=3.0pt, mark=triangle, mark options={solid, mycolor3}]
  table[row sep=crcr]{%
0.345895072941106	0.228793346342776\\
0.398914485462443	0.268885373580139\\
0.447944293311666	0.2644378647174\\
};

\addplot [color=mycolor4, line width=2.0pt, mark size=3.0pt, mark=triangle, mark options={solid, mycolor4}]
  table[row sep=crcr]{%
0.345895072941106	0.241718130039549\\
0.398914485462443	0.279278115605883\\
0.447944293311666	0.263555344761205\\
};

\addplot [color=mycolor5, line width=2.0pt, mark size=3.0pt, mark=triangle, mark options={solid, mycolor5}]
  table[row sep=crcr]{%
0.345895072941106	0.371667212742591\\
0.398914485462443	0.404545984391672\\
0.447944293311666	0.403037398460091\\
};

\addplot [color=mycolor6, line width=2.0pt, mark size=3.0pt, mark=triangle, mark options={solid, mycolor6}]
  table[row sep=crcr]{%
0.345895072941106	0.428345742215812\\
0.398914485462443	0.415569282049536\\
0.447944293311666	0.401814869041727\\
};

\end{axis}

\begin{axis}[%
width=0.951\fwidth,
height=\fheight,
at={(0\fwidth,0\fheight)},
scale only axis,
xmin=0.34,
xmax=0.455,
xtick={0.345895072941106,0.398914485462443,0.447944293311666},
xticklabels={{6000},{5000},{4000}},
xlabel style={font=\color{white!15!black}},
xlabel={Average rotor speed (rpm)},
ymin=0,
ymax=1,
ytick={\empty},
axis x line*=top,
axis y line*=right,
xmajorgrids,
ymajorgrids,
legend style={legend cell align=left, align=left, draw=white!15!black, fill=white!94!black}
]
\end{axis}
\end{tikzpicture}%

%% file: plots/stoi_improv_Ma_8.tex
%
%
\definecolor{mycolor1}{rgb}{0.00000,0.44700,0.74100}%
\definecolor{mycolor2}{rgb}{0.85000,0.32500,0.09800}%
\definecolor{mycolor3}{rgb}{0.92900,0.69400,0.12500}%
\definecolor{mycolor4}{rgb}{0.00000,0.44706,0.74118}%
\definecolor{mycolor5}{rgb}{0.85098,0.32549,0.09804}%
\definecolor{mycolor6}{rgb}{0.92941,0.69412,0.12549}%
\begin{tikzpicture}

\begin{axis}[%
  width=0.951\fwidth,
  height=\fheight,
  at={(0\fwidth,0\fheight)},
  scale only axis,
  xmin=0.34,
  xmax=0.455,
  xlabel style={font=\color{white!15!black}},
  xlabel={Original STOI},
  ymin=0,
  ymax=0.8,
  ylabel style={at={(axis description cs:-.14,.5)},anchor=south,font=\color{white!15!black}},
  ylabel={STOI improvement},
  axis background/.style={fill=white},
  xmajorgrids,
  ymajorgrids,
  xtick={0.345895072941106,0.398914485462443,0.447944293311666},
]
\addplot [color=mycolor1, dashed, line width=2.0pt, mark size=3.0pt, mark=triangle, mark options={solid, mycolor1}]
  table[row sep=crcr]{%
0.345895072941106	0.18578066937493\\
0.398914485462443	0.270602758771189\\
0.447944293311666	0.402118693769455\\
};

\addplot [color=mycolor2, dashed, line width=2.0pt, mark size=3.0pt, mark=triangle, mark options={solid, mycolor2}]
  table[row sep=crcr]{%
0.345895072941106	0.197500728254079\\
0.398914485462443	0.283129820825849\\
0.447944293311666	0.430072156588714\\
};

\addplot [color=mycolor3, dashed, line width=2.0pt, mark size=3.0pt, mark=triangle, mark options={solid, mycolor3}]
  table[row sep=crcr]{%
0.345895072941106	0.298231715690508\\
0.398914485462443	0.381379395344257\\
0.447944293311666	0.42957020144324\\
};

\addplot [color=mycolor4, line width=2.0pt, mark size=3.0pt, mark=triangle, mark options={solid, mycolor4}]
  table[row sep=crcr]{%
0.345895072941106	0.603746243963795\\
0.398914485462443	0.630313216854405\\
0.447944293311666	0.574627976394973\\
};

\addplot [color=mycolor5, line width=2.0pt, mark size=3.0pt, mark=triangle, mark options={solid, mycolor5}]
  table[row sep=crcr]{%
0.345895072941106	0.696721289238499\\
0.398914485462443	0.690926829669284\\
0.447944293311666	0.609264843846353\\
};

\addplot [color=mycolor6, line width=2.0pt, mark size=3.0pt, mark=triangle, mark options={solid, mycolor6}]
  table[row sep=crcr]{%
0.345895072941106	0.726507687779234\\
0.398914485462443	0.698030375317964\\
0.447944293311666	0.610883526067982\\
};

\end{axis}

\begin{axis}[%
width=0.951\fwidth,
height=\fheight,
at={(0\fwidth,0\fheight)},
scale only axis,
xmin=0.34,
xmax=0.455,
xtick={0.345895072941106,0.398914485462443,0.447944293311666},
xticklabels={{6000},{5000},{4000}},
xlabel style={font=\color{white!15!black}},
xlabel={Average rotor speed (rpm)},
ymin=0,
ymax=1,
ytick={\empty},
axis x line*=top,
axis y line*=right,
xmajorgrids,
ymajorgrids,
legend style={legend cell align=left, align=left, draw=white!15!black, fill=white!94!black}
]
\end{axis}
\end{tikzpicture}%

%% file: plots/stoi_improv_Ma_12.tex
%
%
\definecolor{mycolor1}{rgb}{0.00000,0.44700,0.74100}%
\definecolor{mycolor2}{rgb}{0.85000,0.32500,0.09800}%
\definecolor{mycolor3}{rgb}{0.92900,0.69400,0.12500}%
\definecolor{mycolor4}{rgb}{0.00000,0.44706,0.74118}%
\definecolor{mycolor5}{rgb}{0.85098,0.32549,0.09804}%
\definecolor{mycolor6}{rgb}{0.92941,0.69412,0.12549}%
\begin{tikzpicture}

\begin{axis}[%
    width=0.951\fwidth,
    height=\fheight,
    at={(0\fwidth,0\fheight)},
    scale only axis,
    xmin=0.34,
    xmax=0.455,
    xlabel style={font=\color{white!15!black}},
    xlabel={Original STOI},
    ymin=0,
    ymax=0.8,
    ylabel style={at={(axis description cs:-.14,.5)},anchor=south,font=\color{white!15!black}},
    ylabel={STOI improvement},
    axis background/.style={fill=white},
    xmajorgrids,
    ymajorgrids,
    xtick={0.345895072941106,0.398914485462443,0.447944293311666},
]
\addplot [color=mycolor1, dashed, line width=2.0pt, mark size=3.0pt, mark=triangle, mark options={solid, mycolor1}]
  table[row sep=crcr]{%
0.345895072941106	0.18380396011427\\
0.398914485462443	0.284040746984259\\
0.447944293311666	0.415040016428609\\
};

\addplot [color=mycolor2, dashed, line width=2.0pt, mark size=3.0pt, mark=triangle, mark options={solid, mycolor2}]
  table[row sep=crcr]{%
0.345895072941106	0.194158613281734\\
0.398914485462443	0.295533362950885\\
0.447944293311666	0.4376801454839\\
};

\addplot [color=mycolor3, dashed, line width=2.0pt, mark size=3.0pt, mark=triangle, mark options={solid, mycolor3}]
  table[row sep=crcr]{%
0.345895072941106	0.292046517500901\\
0.398914485462443	0.385848647801013\\
0.447944293311666	0.436208201827163\\
};

\addplot [color=mycolor4, line width=2.0pt, mark size=3.0pt, mark=triangle, mark options={solid, mycolor4}]
  table[row sep=crcr]{%
0.345895072941106	0.654582138458538\\
0.398914485462443	0.65704129867521\\
0.447944293311666	0.596901419383758\\
};

\addplot [color=mycolor5, line width=2.0pt, mark size=3.0pt, mark=triangle, mark options={solid, mycolor5}]
  table[row sep=crcr]{%
0.345895072941106	0.724697139535271\\
0.398914485462443	0.711738308433264\\
0.447944293311666	0.627453317433456\\
};

\addplot [color=mycolor6, line width=2.0pt, mark size=3.0pt, mark=triangle, mark options={solid, mycolor6}]
  table[row sep=crcr]{%
0.345895072941106	0.759054244664058\\
0.398914485462443	0.719501694044504\\
0.447944293311666	0.640595364587255\\
};

\end{axis}

\begin{axis}[%
width=0.951\fwidth,
height=\fheight,
at={(0\fwidth,0\fheight)},
scale only axis,
xmin=0.34,
xmax=0.455,
xtick={0.345895072941106,0.398914485462443,0.447944293311666},
xticklabels={{6000},{5000},{4000}},
xlabel style={font=\color{white!15!black}},
xlabel={Average rotor speed (rpm)},
ymin=0,
ymax=1,
ytick={\empty},
axis x line*=top,
axis y line*=right,
xmajorgrids,
ymajorgrids,
legend style={legend cell align=left, align=left, draw=white!15!black, fill=white!94!black}
]
\end{axis}
\end{tikzpicture}%

%% file: plots/pesq_improv_Ma_4.tex
%
%
\definecolor{mycolor1}{rgb}{0.00000,0.44700,0.74100}%
\definecolor{mycolor2}{rgb}{0.85000,0.32500,0.09800}%
\definecolor{mycolor3}{rgb}{0.92900,0.69400,0.12500}%
\definecolor{mycolor4}{rgb}{0.00000,0.44706,0.74118}%
\definecolor{mycolor5}{rgb}{0.85098,0.32549,0.09804}%
\definecolor{mycolor6}{rgb}{0.92941,0.69412,0.12549}%
\begin{tikzpicture}

\begin{axis}[%
width=0.951\fwidth,
height=\fheight,
at={(0\fwidth,0\fheight)},
scale only axis,
xmin=1,
xmax=1.5,
xlabel style={font=\color{white!15!black}},
xlabel={Original PESQ score},
ymin=0,
ymax=0.8,
ylabel style={at={(axis description cs:-.14,.5)},anchor=south,font=\color{white!15!black}},
ylabel={PESQ score improvement},
axis background/.style={fill=white},
ymajorgrids,
legend style={legend cell align=left, align=left, draw=white!15!black}
]
\addplot [color=mycolor1, dashed, line width=2.0pt, mark size=3.0pt, mark=square, mark options={solid, mycolor1}]
  table[row sep=crcr]{%
1.022	0.255381604696673\\
1.334	0.0427286356821588\\
1.485	0.104377104377104\\
};

\addplot [color=mycolor2, dashed, line width=2.0pt, mark size=3.0pt, mark=square, mark options={solid, mycolor2}]
  table[row sep=crcr]{%
1.022	0.299412915851272\\
1.334	0.0809595202398801\\
1.485	0.156228956228956\\
};

\addplot [color=mycolor3, dashed, line width=2.0pt, mark size=3.0pt, mark=square, mark options={solid, mycolor3}]
  table[row sep=crcr]{%
1.022	0.448140900195695\\
1.334	0.204647676161919\\
1.485	0.152861952861953\\
};

\addplot [color=mycolor4, line width=2.0pt, mark size=3.0pt, mark=square, mark options={solid, mycolor4}]
  table[row sep=crcr]{%
1.022	0.413894324853229\\
1.334	0.178410794602699\\
1.485	0.0653198653198652\\
};

\addplot [color=mycolor5, line width=2.0pt, mark size=3.0pt, mark=square, mark options={solid, mycolor5}]
  table[row sep=crcr]{%
1.022	0.410958904109589\\
1.334	0.251124437781109\\
1.485	0.118518518518518\\
};

\addplot [color=mycolor6, line width=2.0pt, mark size=3.0pt, mark=square, mark options={solid, mycolor6}]
  table[row sep=crcr]{%
1.022	0.501956947162427\\
1.334	0.349325337331334\\
1.485	0.268686868686869\\
};

\end{axis}

\begin{axis}[%
width=0.951\fwidth,
height=\fheight,
at={(0\fwidth,0\fheight)},
scale only axis,
xmin=1,
xmax=1.5,
xtick={1.022,1.334,1.485},
xticklabels={{6000},{5000},{4000}},
xlabel style={font=\color{white!15!black}},
xlabel={Average rotor speed (rpm)},
ymin=0,
ymax=1,
ytick={\empty},
axis x line*=top,
axis y line*=right,
xmajorgrids,
ymajorgrids,
legend style={legend cell align=left, align=left, draw=white!15!black, fill=white!94!black}
]
\end{axis}
\end{tikzpicture}%

%% file: plots/pesq_improv_Ma_8.tex
%
%
\definecolor{mycolor1}{rgb}{0.00000,0.44700,0.74100}%
\definecolor{mycolor2}{rgb}{0.85000,0.32500,0.09800}%
\definecolor{mycolor3}{rgb}{0.92900,0.69400,0.12500}%
\definecolor{mycolor4}{rgb}{0.00000,0.44706,0.74118}%
\definecolor{mycolor5}{rgb}{0.85098,0.32549,0.09804}%
\definecolor{mycolor6}{rgb}{0.92941,0.69412,0.12549}%
\begin{tikzpicture}

\begin{axis}[%
width=0.951\fwidth,
height=\fheight,
at={(0\fwidth,0\fheight)},
scale only axis,
xmin=1,
xmax=1.5,
xlabel style={font=\color{white!15!black}},
xlabel={Original PESQ score},
ymin=0.1,
ymax=0.8,
ylabel style={at={(axis description cs:-.14,.5)},anchor=south,font=\color{white!15!black}},
ylabel={PESQ score improvement},
axis background/.style={fill=white},
ymajorgrids,
legend style={legend cell align=left, align=left, draw=white!15!black}
]
\addplot [color=mycolor1, dashed, line width=2.0pt, mark size=3.0pt, mark=square, mark options={solid, mycolor1}]
  table[row sep=crcr]{%
1.022	0.341487279843444\\
1.334	0.16191904047976\\
1.485	0.303030303030303\\
};

\addplot [color=mycolor2, dashed, line width=2.0pt, mark size=3.0pt, mark=square, mark options={solid, mycolor2}]
  table[row sep=crcr]{%
1.022	0.36399217221135\\
1.334	0.183658170914543\\
1.485	0.315151515151515\\
};

\addplot [color=mycolor3, dashed, line width=2.0pt, mark size=3.0pt, mark=square, mark options={solid, mycolor3}]
  table[row sep=crcr]{%
1.022	0.531311154598826\\
1.334	0.326836581709145\\
1.485	0.326599326599327\\
};

\addplot [color=mycolor4, line width=2.0pt, mark size=3.0pt, mark=square, mark options={solid, mycolor4}]
  table[row sep=crcr]{%
1.022	0.674168297455969\\
1.334	0.467766116941529\\
1.485	0.432323232323232\\
};

\addplot [color=mycolor5, line width=2.0pt, mark size=3.0pt, mark=square, mark options={solid, mycolor5}]
  table[row sep=crcr]{%
1.022	0.74853228962818\\
1.334	0.511244377811094\\
1.485	0.465993265993266\\
};

\addplot [color=mycolor6, line width=2.0pt, mark size=3.0pt, mark=square, mark options={solid, mycolor6}]
  table[row sep=crcr]{%
1.022	0.76320939334638\\
1.334	0.583208395802099\\
1.485	0.552861952861953\\
};

\end{axis}

\begin{axis}[%
width=0.951\fwidth,
height=\fheight,
at={(0\fwidth,0\fheight)},
scale only axis,
xmin=1,
xmax=1.5,
xtick={1.022,1.334,1.485},
xticklabels={{6000},{5000},{4000}},
xlabel style={font=\color{white!15!black}},
xlabel={Average rotor speed (rpm)},
ymin=0,
ymax=0.8,
ytick={\empty},
axis x line*=top,
axis y line*=right,
xmajorgrids,
ymajorgrids,
legend style={legend cell align=left, align=left, draw=white!15!black, fill=white!94!black}
]
\end{axis}
\end{tikzpicture}%

%% file: plots/pesq_improv_Ma_12.tex
%
%
\definecolor{mycolor1}{rgb}{0.00000,0.44700,0.74100}%
\definecolor{mycolor2}{rgb}{0.85000,0.32500,0.09800}%
\definecolor{mycolor3}{rgb}{0.92900,0.69400,0.12500}%
\definecolor{mycolor4}{rgb}{0.00000,0.44706,0.74118}%
\definecolor{mycolor5}{rgb}{0.85098,0.32549,0.09804}%
\definecolor{mycolor6}{rgb}{0.92941,0.69412,0.12549}%
\begin{tikzpicture}

\begin{axis}[%
width=0.951\fwidth,
height=\fheight,
at={(0\fwidth,0\fheight)},
scale only axis,
xmin=1,
xmax=1.5,
xlabel style={font=\color{white!15!black}},
xlabel={Original PESQ score},
ymin=0.1,
ymax=0.8,
ylabel style={at={(axis description cs:-.14,.5)},anchor=south,font=\color{white!15!black}},
ylabel={PESQ score improvement},
axis background/.style={fill=white},
ymajorgrids,
legend style={legend cell align=left, align=left, draw=white!15!black}
]
\addplot [color=mycolor1, dashed, line width=2.0pt, mark size=3.0pt, mark=square, mark options={solid, mycolor1}]
  table[row sep=crcr]{%
1.022	0.388454011741683\\
1.334	0.18215892053973\\
1.485	0.307070707070707\\
};

\addplot [color=mycolor2, dashed, line width=2.0pt, mark size=3.0pt, mark=square, mark options={solid, mycolor2}]
  table[row sep=crcr]{%
1.022	0.347358121330724\\
1.334	0.197151424287856\\
1.485	0.317845117845118\\
};

\addplot [color=mycolor3, dashed, line width=2.0pt, mark size=3.0pt, mark=square, mark options={solid, mycolor3}]
  table[row sep=crcr]{%
1.022	0.538160469667319\\
1.334	0.330584707646177\\
1.485	0.330639730639731\\
};

\addplot [color=mycolor4, line width=2.0pt, mark size=3.0pt, mark=square, mark options={solid, mycolor4}]
  table[row sep=crcr]{%
1.022	0.703522504892368\\
1.334	0.500749625187406\\
1.485	0.463973063973064\\
};

\addplot [color=mycolor5, line width=2.0pt, mark size=3.0pt, mark=square, mark options={solid, mycolor5}]
  table[row sep=crcr]{%
1.022	0.756360078277886\\
1.334	0.535232383808096\\
1.485	0.505723905723906\\
};

\addplot [color=mycolor6, line width=2.0pt, mark size=3.0pt, mark=square, mark options={solid, mycolor6}]
  table[row sep=crcr]{%
1.022	0.772015655577299\\
1.334	0.608695652173913\\
1.485	0.594612794612794\\
};

\end{axis}

\begin{axis}[%
width=0.951\fwidth,
height=\fheight,
at={(0\fwidth,0\fheight)},
scale only axis,
xmin=1,
xmax=1.5,
xtick={1.022,1.334,1.485},
xticklabels={{6000},{5000},{4000}},
xlabel style={font=\color{white!15!black}},
xlabel={Average rotor speed (rpm)},
ymin=0,
ymax=0.8,
ytick={\empty},
axis x line*=top,
axis y line*=right,
xmajorgrids,
ymajorgrids,
legend style={legend cell align=left, align=left, draw=white!15!black, fill=white!94!black}
]
\end{axis}
\end{tikzpicture}%

%% file: ICA2022_ABS-0237.bbl
\noindent \section*{\uppercase{References}}
\vspace{-18pt}

\begin{thebibliography}{5}


\bibitem{dregon}
Strauss M, Mordel P, Miguet V, Deleforge A. DREGON: Dataset and methods for UAV-embedded sound source localization. In: 2018 IEEE/RSJ International Conference on Intelligent Robots and Systems (IROS). Madrid: IEEE; 2018. p. 1–8.

\bibitem{hioka_speechenhancement}
Hioka Y, Kingan M, Schmid G, Stol KA. Speech enhancement using a microphone array mounted on an unmanned aerial vehicle. In: 2016 IEEE International Workshop on Acoustic Signal Enhancement (IWAENC). Xi’an, China: IEEE; 2016. p. 1–5.

\bibitem{nasa}
Hubbard HH, editor. Aeroacoustics of flight vehicles: theory and practice. NASA Office of Management, Scientific and Technical Information Program; 1991.

\bibitem{book_hioka}
Hioka Y, Yen B, McKay R, Kingan M. Clean audio recording using unmanned aerial vehicles. In: Unmanned Aerial Systems. Elsevier; 2021. p. 175–202.

\bibitem{gannot}
Gannot S, Vincent E, Markovich-Golan S, Ozerov A. A consolidated perspective on multimicrophone speech enhancement and source separation. IEEE/ACM Trans Audio Speech Lang Process. 2017 Apr;25(4):692–730.

\bibitem{cavallaro}
Wang L, Cavallaro A. A blind source separation framework for ego-noise reduction on multi-rotor drones. IEEE/ACM Trans Audio Speech Lang Process. 2020;28:2523–37.

\bibitem{moonen}
Doclo S, Moonen M. GSVD-based optimal filtering for single and multimicrophone speech enhancement. IEEE Trans Signal Process. 2002 Sep;50(9):2230–44.

\bibitem{ngo}
Ngo K, Spriet A, Moonen M, Wouters J, Jensen SH. Incorporating the conditional speech presence probability in multi-channel Wiener filter based noise reduction in hearing aids. EURASIP J Adv Signal Process. 2009 Dec;2009(1):930625.

\bibitem{rompaey}
Rompaey RVan, Moonen M. Distributed adaptive node-specific signal estimation in a wireless sensor network with partial prior knowledge of the desired source steering vector. In: 2019 27th European Signal Processing Conference (EUSIPCO). A Coruna, Spain: 2019. p. 1–5.

\bibitem{github}
Tengan E, Dietzen T, Ruiz S, Alkmim M, Cardenuto J, van Waterschoot T. Replication data for: Speech enhancement using ego-noise references with a microphone array embedded in an unmanned aerial vehicle. KU Leuven RDR; 2022. Available from: \url{https://doi.org/10.48804/PZAVUC}.

\bibitem{serizel}
Serizel R, Moonen M, Van Dijk B, Wouters J. Low-rank approximation based multichannel Wiener filter algorithms for noise reduction with application in cochlear implants. IEEE/ACM Trans Audio Speech Lang Process. 2014 Apr;22(4):785–99.

\bibitem{strauss}
Breed BR, Strauss J. A short proof of the equivalence of LCMV and GSC beamforming. IEEE Signal Process Lett. 2002 Jun;9(6):168–9.

\bibitem{hover}
Hioka Y, Kingan M, Schmid G, McKay R, Stol KA. Design of an unmanned aerial vehicle mounted system for quiet audio recording. Applied Acoustics. 2019 Dec;155:423–7.

\bibitem{corpus}
Yamagishi J, Veaux C, MacDonald K. {CSTR} {VCTK} corpus: English multi-speaker corpus for {CSTR} voice cloning toolkit (version 0.92). University of Edinburgh. The Centre for Speech Technology Research. 2019.

\bibitem{spp}
Gerkmann T, Krawczyk M, Martin R. Speech presence probability estimation based on temporal cepstrum smoothing. In: 2010 IEEE International Conference on Acoustics, Speech and Signal Processing (ICASSP)). Dallas, TX, USA: IEEE; 2010. p. 4254–7.

\bibitem{stoi}
Taal CH, Hendriks RC, Heusdens R, Jensen J. An algorithm for intelligibility prediction of time–frequency weighted noisy speech. Vol. 19, IEEE/ACM Trans Audio Speech Lang Process. 2011 Sep;19(7):2125–36.

\bibitem{pesq}
Rix AW, Beerends JG, Hollier MP, Hekstra AP. Perceptual evaluation of speech quality (PESQ) - a new method for speech quality assessment of telephone networks and codecs. In: 2001 IEEE International Conference on Acoustics, Speech, and Signal Processing Proceedings (ICASSP). Salt Lake City, UT, USA: IEEE; 2001. p. 749–52.

\end{thebibliography}
